\documentclass[lettersize,journal]{IEEEtran}
\usepackage{amsmath,amsfonts}
\usepackage{algorithmic}
\usepackage{algorithm}
\usepackage{array}
\usepackage[caption=false,font=normalsize,labelfont=sf,textfont=sf]{subfig}
\usepackage{textcomp}
\usepackage{stfloats}
\usepackage{url}
\usepackage{verbatim}
\usepackage{graphicx}
\usepackage{cite}
\usepackage{caption2}
\usepackage{epsfig,lipsum,tabularx,array,adjustbox,subfig, enumitem, algorithm, algorithmic}
\begin{document}

\title{MeloTrans: A Text to Symbolic Music Generation Model Following Human Composition Habit}

\author{Yutian Wang,~\IEEEmembership{Member,~IEEE}, Wanyin Yang, Zhenrong Dai, Yilong Zhang, Kun Zhao, Hui Wang, ~\IEEEmembership{Member,~IEEE}
\thanks{This work was supported by National Key Research and Development Project of China (2022YFF0902402)}
\thanks{Yutian Wang and Wanyin Yang are co-first authors. Hui Wang is corresponding author.}
\thanks{Yutian Wang, Wanyin Yang, Zhenrong Dai are with Key Laboratory of Media Audio $\&$ Video (Communication University of China), Ministry of Education. Communication University of China, Beijing 100024. (e-mail: wangyutian@cuc.edu.cn; ywanyin@foxmail.com; dzr9696@163.com)\\ Yilong Zhang is with School of Music and Recording Arts, Communication University of China, Beijing 100024. (e-mail: zhangyilong@cuc.edu.cn)\\ Kun Zhao and Hui Wang are with State Key Laboratory of Media Convergence and Communication, Communication University of China, Beijing 100024. (e-mail: zhaokun\_213@cuc.edu.cn; hwang@cuc.edu.cn)}}

\markboth{Journal of \LaTeX\ Class Files,~Vol.~14, No.~8, August~2021}%
{Shell \MakeLowercase{\textit{Y. Wang et al.}}: MeloTrans: A Text to Symbolic Music Generation Model Following Human Composition Habit}


\maketitle

\begin{abstract}
At present, neural network models show powerful sequence prediction ability and are used in many automatic composition models. In comparison, the way humans compose music is very different from it. Composers usually start by creating musical motifs and then develop them into music through a series of rules. This process ensures that the music has a specific structure and changing pattern. However, it is difficult for neural network models to learn these composition rules from training data, which results in a lack of musicality and diversity in the generated music. This paper posits that integrating the learning capabilities of neural networks with human-derived knowledge may lead to better results. To archive this, we develop the POP909$\_$M dataset, the first to include labels for musical motifs and their variants, providing a basis for mimicking human compositional habits. Building on this, we propose MeloTrans, a text-to-music composition model that employs principles of motif development rules. Our experiments demonstrate that MeloTrans excels beyond existing music generation models and even surpasses Large Language Models (LLMs) like ChatGPT-4. This highlights the importance of merging human insights with neural network capabilities to achieve superior symbolic music generation.
\end{abstract}


\section{Introduction}
\IEEEPARstart{S}{ymbolic} music generation have been a subject of extensive research in recent years. Transformer-based models have shown good performance in modeling long sequences, thus been widely employed in symbolic music generation \cite{roberts_hierarchical_2018}\cite{huang_music_2018}\cite{hsiao_compound_2021}. However, these models often lack a sense of musicality. For instance, human-composed music frequently features motif repetition and deformation. 
\begin{figure}[htb]
	\centering
	\subfloat[]{
		\includegraphics[width=3.5cm,height=6cm]{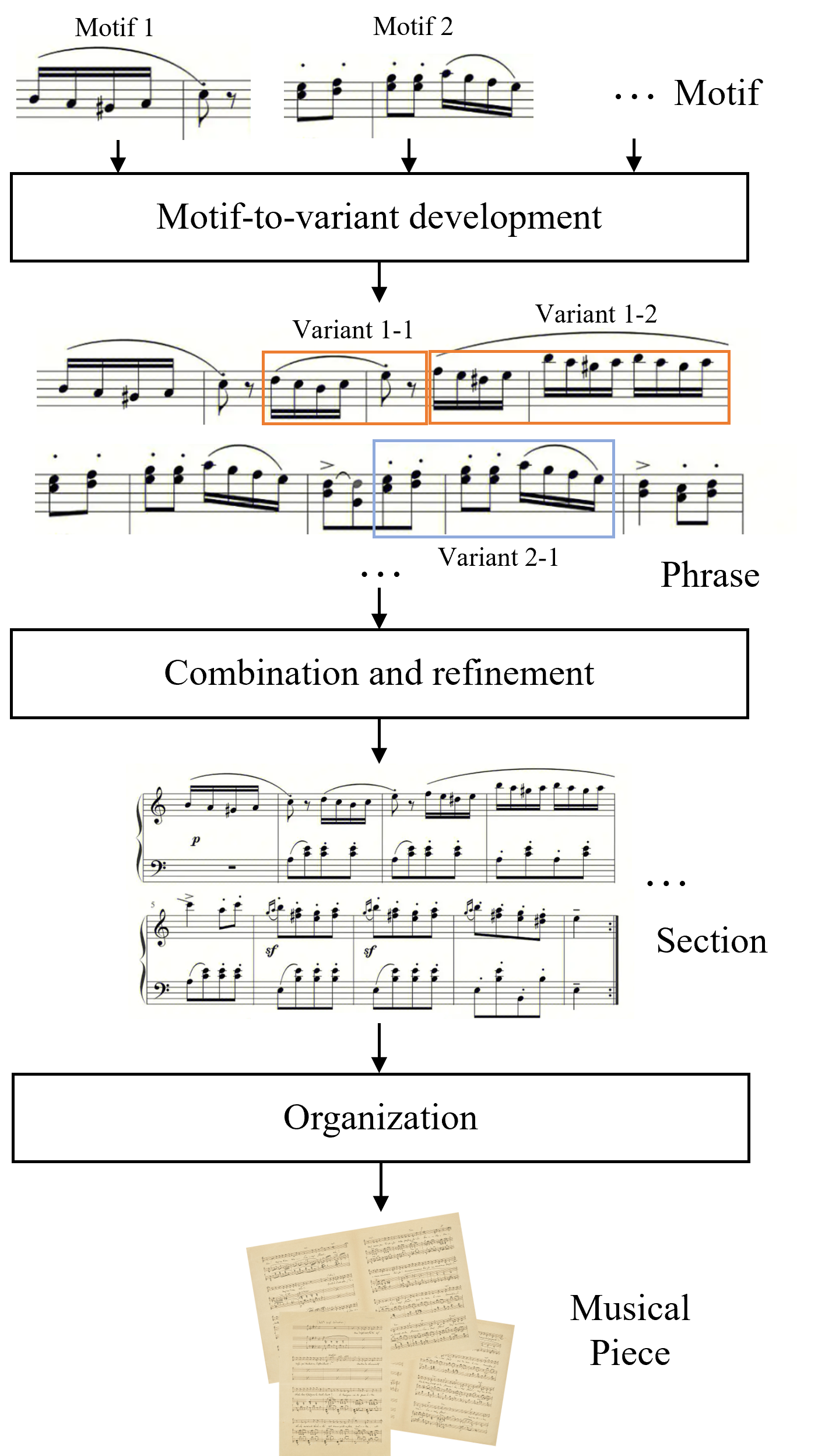}
	}
	\hfil
	\subfloat[]{
		\includegraphics[width=3.5cm,height=6cm]{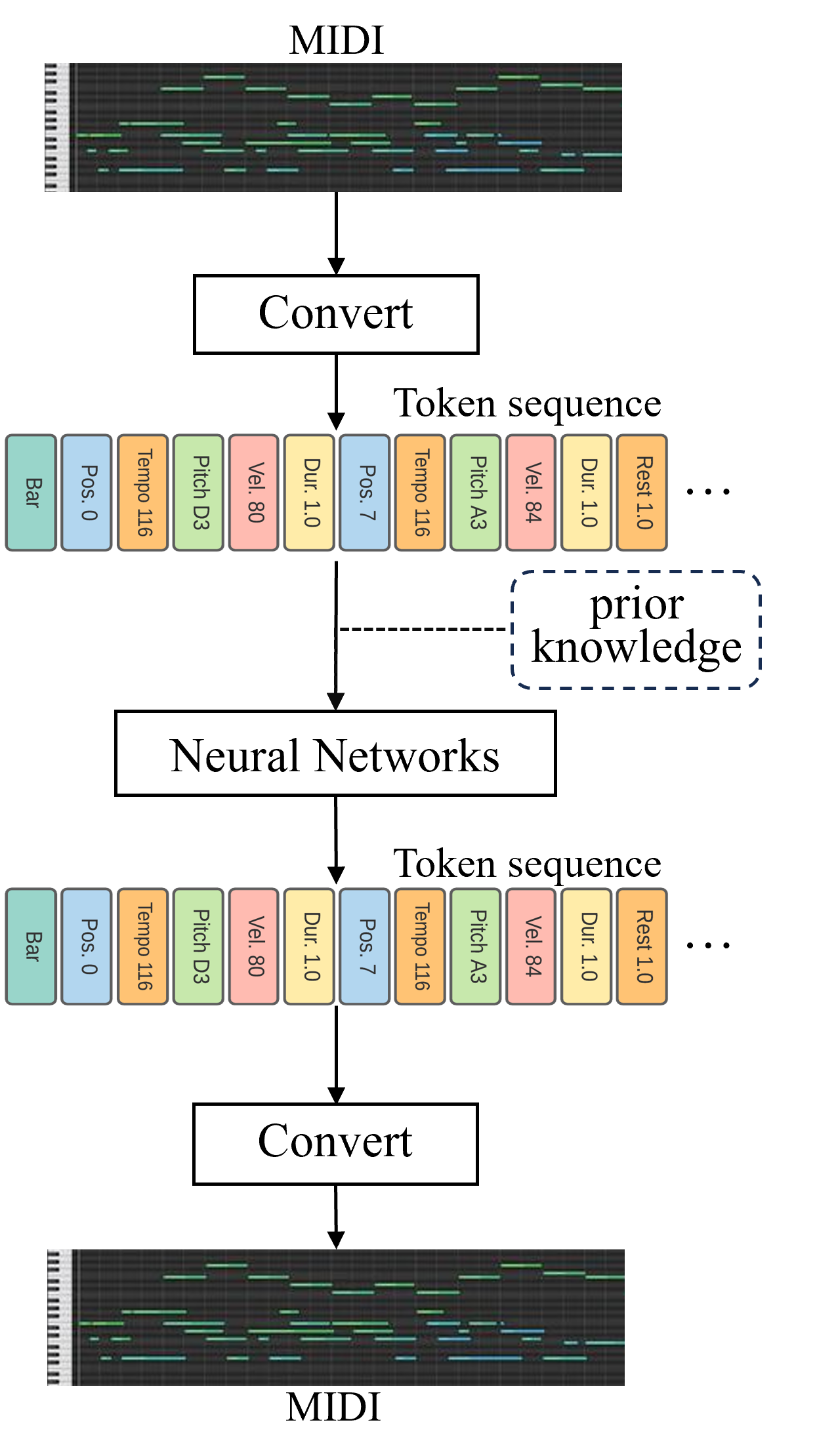}
	}
	
	\caption{Comparison of \textbf{(a)} human composition process and \textbf{(b)} neural network model composition process.}
	\label{comparison_of_human_and_machine}
\end{figure}
Such structural elements tend to be absent in output from neural network composition models, primarily due to the difficulty in extracting such complex patterns from training data.
Numerous studies have attempted to incorporate prior knowledge into the generation process. Considering repetition of motifs as a significant musical pattern, some works strive to emulate this by means of detecting musical phrases and sections \cite{zhao_accomontage_2021}\cite{shih_theme_2023}\cite{yi_accomontage2_2022}. However, the integration of repetition mode in these works often results in an unnatural sound, appearing as if it's repetition for repetition's sake. Other research have concentrated on learning relationships between bars, achieving structural coherence based on bar similarity \cite{wu_popmnet_2020}\cite{zou_melons_2022}\cite{wu_power_2023}. Nonetheless, these relationships do not fully represent the essence of musical structure. Further efforts combine expert systems with neural networks but continue to struggle with producing monotonous music due to prescriptive rules\cite{lu_meloform_2022}. Despite considering prior knowledge, these works fail to align with human composition habits, resulting in music that lacks diversity and musicality. 

Figure \ref{comparison_of_human_and_machine} illustrates the different idea of human and machine composition. In human composition, the process typically initiates with the creation of motifs, subsequently varies and develops into music  phrases \cite{kachulis_songwriters_2003} \cite{kidde_learning_2020} \cite{hernandez-olivan_music_2023}. These phrases undergo adjustment and integration, conforming to the composition rules, thereby culminating in a complete music segment. The motif's development is termed as 'variant'. In contrast, machine composition is essentially a conditional sequential prediction model, which is hard to learn all of the internal rules of music. Such models often come at the cost of poor performance or a large number of model parameters. Therefore, introduce the human composition experience into neural networks is crucial for automatic composition models.

Two primary challenges exist in aligning neural networks with the human composition paradigm: 1) generating motifs and 2) developing motifs into musical pieces. In this study, we introduce the POP909\_M dataset, which annotates motifs, variants, and textual descriptions of music from the POP909 dataset \cite{wang_pop909_2020}. Building upon this dataset, we propose MeloTrans, a two-stage text-to-music generative model. The first stage involves text-to-motif generation. This multi-modal generation process usually involves complex alignment challenges. Given the typically short length of motifs in our system, we designed a Valence and Arousal based generation algorithm, which greatly reduces the system complexity while maintaining the generation performance. The second stage encompasses motif-to-melody transformation. We developed a multi-branch model that simulates human composition rules, ensuring the synthesized melody exhibits a clear structure. By introducing a motif-driven composition pipeline, our approach facilitates the generation of more expressive and coherent musical compositions. The main contributions of this paper are as follows:
\begin{itemize}[itemsep=0pt,topsep=2pt,parsep=0pt,partopsep=0pt]
	\item We propose POP909\_M dataset, labeling motifs, 5 types of variants and the text descriptions on all the songs. To our knowledge, this is the first dataset to label motif and variants.
	\item A multi-branch model is adopted to generate different variants, which can guarantee the musical structure of melody.
	\item A positional encoding method aligned with multiple variants enables the model to effectively learn various motif-variant development rules.
	\item Experiments show that our model outperforms state-of-the-art models trained with massive amounts of data and large language models in terms of structural and semantic matchability.
\end{itemize}
The rest of this paper is structured as follows: Section 2 describes the related works. Section 3 demonstrate the POP909\_M dataset with different aspect. Section 4 is the proposed MeloTrans model with descriptions of modules. Section 5 presents the experiments and settings and the results and discussions. Conclusions are drawn in Section 6.

\section{Related Works}

Machine composition has two synthetic targets, the first is to directly synthesize musical waveform, and the second is to synthesize symbolic music. Both have made great progress in recent years. For the former, it is mainly realized by large retrieval model \cite{Inc.} or large multimodal model \cite{Riffusion}. The main difficulty in implementing these models is the lack of text-music pairing data, which makes the creation of corresponding datasets an important part of their work \cite{Schneider2023MosaiTG} \cite{Zhu2023ERNIEMusicTM}. To overcome this problem, pre-trained models \cite{Huang2022MuLanAJ}, semi-supervised \cite{Agostinelli2023MusicLMGM} and unsupervised \cite{NEURIPS2023_94b472a1} training methods have been used in composition models.

However, the challenge of generating music waveforms from text lies in their limited editability. Users cannot directly edit the generated music and must continuously adjust the text prompt to regenerate the music, imposing limitations on the composition process. Moreover, certain professional requirements are challenging to implement in waveform generation, such as when users provide motifs and request the model to develop them into musical phrases. In contrast, symbolic music, represented as a series of musical symbols, can be readily interpreted and manipulated by both humans and machines.

Early symbolic music generation models primarily utilized Recurrent Neural Networks (RNNs), Long Short-Term Memory (LSTM) networks, and similar architectures. In 2002, Eck et al. \cite{Eck2002} pioneered the application of LSTM to blues music composition, demonstrating the ability to improvise well-paced and well-structured blues music from short musical inputs. MelodyRNN \cite{Waite2016}, proposed by the Magenta team, improved the capacity to learn long-term structures through the implementation of Lookback RNN and Attention RNN architectures. Owing to the limitations of recurrent neural networks in generating extended sequences, researchers have explored alternative generative models, including Variational Autoencoders (VAEs) and Generative Adversarial Networks (GANs). MusicVAE \cite{roberts_hierarchical_2018} employs a hierarchical decoder for modeling long structured sequences and utilizes a bi-directional RNN as an encoder, enhancing interpolation and reconstruction capabilities. GrooVAE \cite{Gillick2019LearningTG}, an extension of MusicVAE, generates drum melodies by training on a corpus of electronic drum recordings from live performances. Yang et al. \cite{Yang2017MidiNetAC} proposed MidiNet, a GAN-based model that enhances the innovation in Jazz track generation by incorporating previously generated melody and chord information as conditional input to the generator's intermediate convolutional layer, thereby constraining the generation of note types. JazzGAN \cite{Trieu2018JazzGANI} presents a GAN-based monophonic Jazz generative model, utilizing LSTM to improvise over chord progressions and introducing several metrics for evaluating Jazz-specific musical features. MuseGAN, proposed by Dong et al. \cite{Dong2017MuseGANMS}, is widely recognized as the pioneering model capable of generating multi-track polyphonic music.

Capturing long-term dependencies in music has always been a significant challenge for deep learning models. Several studies have proposed different solutions to address this issue\cite{roberts_hierarchical_2018}\cite{waite_generating_2016}. Music Transformer introduced the first Transformer model for symbolic music generation, demonstrating the capacity of Transformer models to generate coherent, minutes-long polyphonic piano music\cite{huang_music_2018}. Since then, various Transformer-based models have been proposed. However, deep learning models, including Transformers, still face difficulty in generating musical phrases and fragments with clear repetition and development. Another approach is to create new representation of music to shorten the input sequence length, but this cannot fundamentally solve the problem\cite{huang_pop_2020}\cite{liu_symphony_2022}\cite{zhang_sdmuse_2023}. Meanwhile, these works tend to use the first few notes of existing fragments as a prompt, without attributing specific meaning to them. 

To solve this problem, researchers have considered introducing prior knowledge to the generating process. Theme Transformer proposed gated parallel attention mechanisms and theme-aligned positional encoding to conditionally reference thematic material in the music generation process\cite{shih_theme_2023}. However, it merely repeats theme fragments without considering the development relationships between these hierarchical structures. MELONS summarizes the relationships between bars into eight types and proposes a two-step task: structure generation and structure-conditioned melody generation\cite{zou_melons_2022}. This approach can generate complete songs with clear long-term structures, but lacks musical richness compared to real music. MeloForm employs an expert system for rule-based melody generation and a Transformer for refinement\cite{lu_meloform_2022}, performing high accuracy in musical form, but the result sounds similar when generated from the same motif. 

Despite significant advancements in symbolic music composition, relatively few studies have focused on generating symbolic music from textual prompts. TransProse \cite{davis-mohammad-2014-generating}, proposed in 2014, established a set of mapping rules for generating music based on the density of emotional words in a given text. However, this approach fails to capture non-emotional textual information, and its performance is constrained by the hand-crafted nature of the mapping rules. Rangarajan et al. \cite{Rangarajan2015} developed three strategies for mapping text to musical notes; however, due to the character-level functionality, the resulting music lacks coherence and fails to capture the semantic content of the text. BUTTER \cite{zhang-etal-2020-butter}, based on the Gated Recurrent Unit (GRU), enables the search and generation of music clips based on specific textual descriptions. While BUTTER employs a data-driven approach, its flexibility is constrained by the use of synthesized texts from specified keywords (i.e., 25 keys, 6 beat types, and 3 styles), limiting its output to 4-bar music clips in ABC format. Wu et al. \cite{Wu2023} introduced a text-music dataset in 2022 and utilized a pre-trained large language model (LLM) for text-controlled music generation, demonstrating the viability of extracting textual features for controlled music generation. Despite fine-tuning the LLM on over 200,000 texts and ABC notated music samples, the model struggles to effectively align sparse musical information in the text with the generated music. MuseCoco \cite{Lu2023} employs musical attributes as an intermediary to generate symbolic music from textual descriptions. This approach enables self-supervised training of the model using substantial amounts of unlabeled data. To address the challenge of limited labeled data, the text-to-attribute understanding stage employs templates and language models such as ChatGPT to synthesize paired text-attribute data.

Instructed with musical input prompts, the large language model ChatGPT-4 \cite{openai2024gpt4} can also generate music from a given motif, but its development is monotonous\cite{bubeck_sparks_2023}. 

Although considering prior knowledge, previous works still fail to touch the essential element in human composition, consequently lack the desired musicality. In this paper, we propose MeloTrans, which can generate music following 5 kinds of motif-to-variant development to improve both humanity and music richness. 

\section{POP909\_M Dataset}

In the field of symbolic music generation, there is no available dataset that includes motifs and their variants. Therefore, in this paper we propose POP909\_M dataset, featuring MIDI files labeled with motifs, corresponding variants and the text description of the motifs, developed from the POP909 dataset\cite{wang_pop909_2020}.

\subsection{Motifs labeling}
For the convenience of labeling, the bridge tracks and piano accompaniment tracks are removed, retaining only the melody tracks. Chord tracks are added to the MIDI files based on the chord label information from POP909, which follows the data organization method of MELONS\cite{zou_melons_2022}. 

As different people may have different understandings of motif, the annotations are made following a given judgment criterion for recognizing motif. The criteria are based on existing materials \cite{Hankun2006} and summarized as follows: 
\begin{enumerate}[itemsep=0pt,topsep=3pt,parsep=0pt, partopsep=0pt]
	\item Including one or more stressed notes;
	\item Including two or more notes;
	\item The size and length are generally within one bar, not more than two bars;
	\item The note group will develop in subsequent music.
\end{enumerate}

To reduce the difficulty and improve the accuracy of labeling, we did not label the repeated motifs in the dataset. The same motif in a MIDI file only needed to be labeled once, for the other ones can be automatically labeled by detecting same note sequences through a sliding candidate window. Two professional composers were employed to perform cross-validation to ensure that the label was done to a consistent standard.

\subsection{Variant Labeling}
Composition theory literature \cite{Hankun2006}\cite{schoenberg_fundamentals_nodate}\cite{mcnamee_review_1986} summarizes the variants into 5 types: Repetition, Progression, Transformation, Expansion/Compression, and Inversion, which are defined in Table \ref{Variant_types}.

\begin{table*}[hbt]
	\renewcommand{\arraystretch}{1.3}
	\caption{Variant types and corresponding description.}
	\label{Variant_types}
	\centering
	\begin{tabularx}{\textwidth} {c c >{\raggedright\arraybackslash}X}
		\hline
		\bfseries Number& \bfseries Type& \bfseries Description\\
		\hline
		1& Repetition& Complete repetition without changing intervals, rhythms, and pitches.
		\\
		2& Progression& Repetition at different pitches, maintaining general consistency in pitch and rhythm.\\
		3& Transformation&Preserving the basic outline of the motif, the changes are more free.
		\\
		4& Expansion/Compression&Expansion within the motif or reducing the phrase by removing secondary notes.
		\\
		5& Inversion& Proceeding in the opposite direction to the original motif material.
		\\
		\hline
	\end{tabularx}
\end{table*}
Based on the established rules and labeled motifs, an algorithm is designed to automatically identify and label the corresponding musical variants. Specifically, for a musical segment containing a motif, the algorithm slides a candidate window across the note sequence and compiles the start-time list of the notes within the window. If the start-time list of the notes within the window matches that of the motif sequence, the algorithm considers the note sequence within the current window to be a variant. The type of the identified variant is determined by the Pitch Match Ratio (PMR) and the Pitch Trend Match Ratio (TMR), where the pitch trend indicates whether the pitch intervals between the current note and the next note are positive, negative, or neutral. The thresholds for TMR and PMR are determined through the observation of data and the professional composers. The detailed variant labeling process is outlined in Algorithm \ref{Variant_Labeling}.

\begin{algorithm}[hbt]
	\caption{Variant Labeling}
	\label{Variant_Labeling}
	\renewcommand{\algorithmicrequire}{ \textbf{Define:}} 
	\renewcommand{\algorithmicensure}{ \textbf{Output:}} 
	\begin{algorithmic}[1]
		
		\STATE $st\_m \gets$ note $start\_time$ list of the motif; 
		\STATE $pitch\_m \gets$ pitch list of the motif; \\
		\STATE $trend\_m \gets$ pitch trend list of the motif; 
		\STATE $win\_start \gets$ end of the motif; 
		\STATE $win\_len \gets$ motif length; 
		\STATE $step \gets$ bar length.
		\WHILE {$win\_start+win\_len \leq$ end time of the fragment}
		\STATE $st\_c \gets$ note $start\_time$ list of the current window.
		\STATE $pitch\_m \gets$ pitch list of the motif.
		\STATE $trend\_m \gets$ pitch trend list of the motif.
		\IF{$st\_m == st\_c$} 
		\STATE $PMR \gets$ proportion of the same items in $pitch\_m$ and $pitch\_c$.
		\STATE $TMR \gets$ proportion of the same items in $trend\_m$ and $trend\_c$.
		\IF{$TMR \geq 0.6$} 
		\IF{$PMR \geq 0.6$} 
		\STATE $variant \ type \gets 1$
		\ELSE
		\STATE $variant \ type \gets 2$
		\ENDIF
		\ELSIF{$0.2 \leq TMR < 0.6$}
		\STATE $variant \ type \gets 3$
		\ELSE
		\STATE $variant \ type \gets 5$
		\ENDIF 
		\IF{$ (st\_m \subseteq st\_c \ \mathbf{and} \ trend\_m \subseteq trend\_c) \newline \mathbf{or} \ (st\_m \supseteq st\_c \ \mathbf{and} \ trend\_m \supseteq trend\_c)$} 
		\STATE $variant \ type \gets 4$
		\ENDIF 
		\ENDIF 
		\STATE $win\_start \ += step$
		\ENDWHILE 
	\end{algorithmic}
\end{algorithm}

\begin{figure*}[hbt]
	\centering
	\centerline{\epsfig{figure=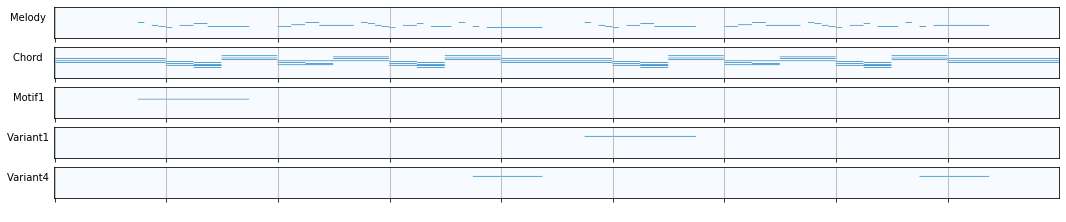,width=18cm}}
	\caption{Demo of data organization in POP909\_M. In the motif and variants tracks, the motif and variants in the corresponding melody track are marked with horizontal lines.}
	\label{fig:data_organization}
\end{figure*}

As illustrated in Figure \ref{fig:data_organization}, following the labeling process, the songs are segmented into 4,419 clips, each comprising musical tracks (i.e., a melody track and a chord track) and label tracks (i.e., a motif label track and 1–5 corresponding variant label tracks), all stored in MIDI format. The songs uniformly adhere to a 4/4 time signature. The dataset encompasses 4,419 motifs and 12,474 variants derived from 860 songs, categorized as follows: 2,744 Repetitions, 1,497 Progressions, 1,372 Transformations, 6,362 Expansions/Compressions, and 499 Inversions.

\subsection{Text description}
One of the primary challenges in the field of text-controlled music generation is the scarcity of text-symbol pair data. Conversely, single-modality symbolic music datasets are readily available. Consequently, this study aims to augment the motifs with text description data to advance research on algorithms related to text and symbolic music.

Due to limited specialized knowledge, individuals who have not undertaken systematic study of music theory are often unable to provide comprehensive descriptions of harmony and musical form, nor can they accurately identify the instruments, rhythmic patterns, and styles employed in musical compositions. For the majority of casual listeners, the most salient aspect of their song descriptions pertains to the evocation of emotion, complemented by contextual applicability and lyrical assessment.

We use a web crawler to extract corresponding playlist names and descriptions of each songs in the POP909\_M dataset from the music website$\footnote{https://music.163.com}$. The data cleaning process encompassed the removal of extreme values, elimination of duplicate descriptions, excision of irrelevant information, and subsequent manual verification. Finally, we have 5,412 playlist data and 12,422 song descriptions.
%

\section{Method}

In this section, we introduce MeloTrans, comprising two modules: Text-to-Motifs Module (TTMM) and Melody Generation Module (MGM). The TTMM is an rule based generation model that takes the input text and turns it into motifs. The MGM is responsible for producing diverse variant types from input motifs and turn them into melodies. 

\subsection{ Text-to-Motifs Module (TTMM)}

\begin{figure}[hbt]
	\centering
	\centerline{\epsfig{figure=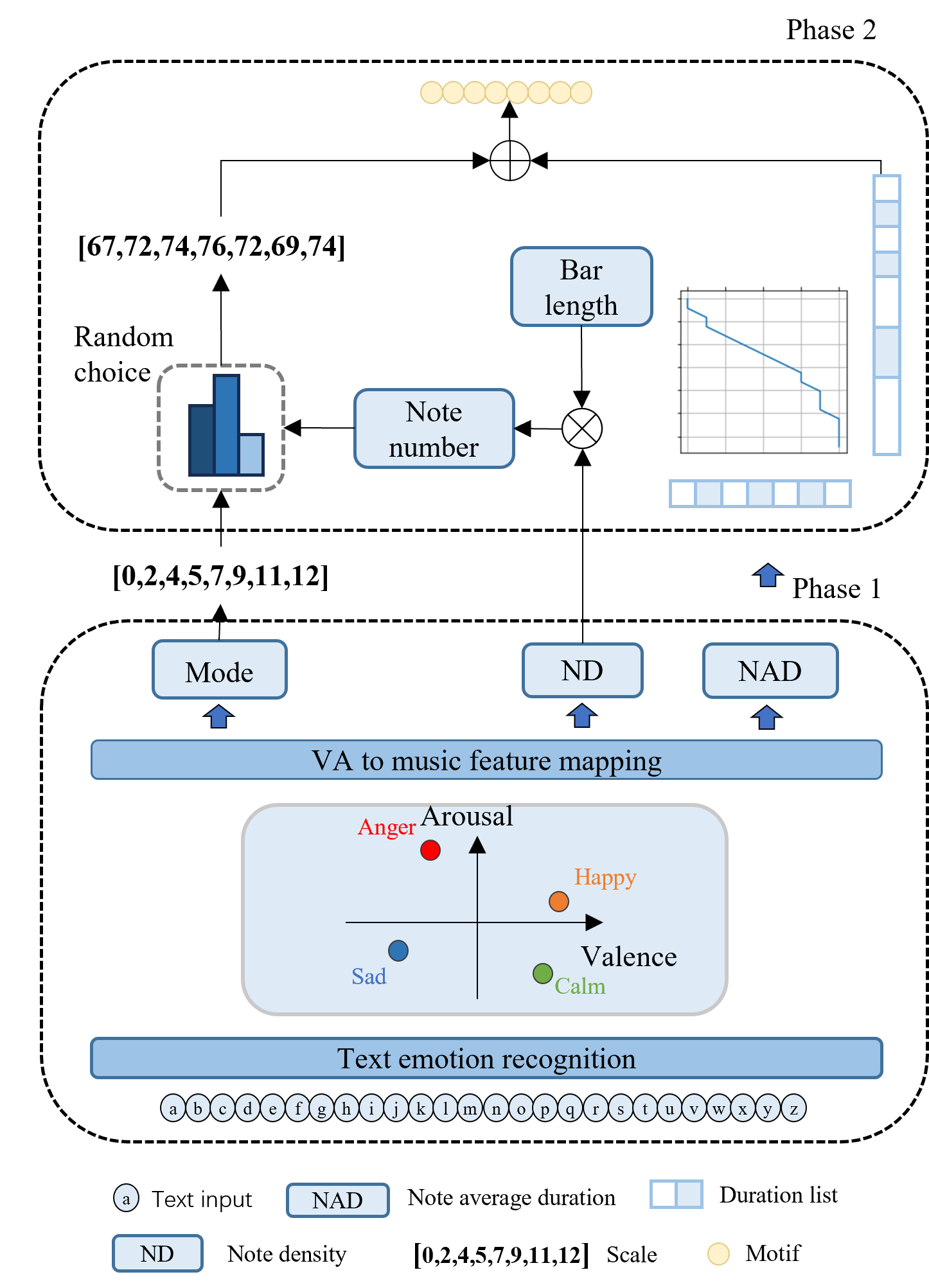,width=8.5cm}}
	\caption{Architecture of TTMM. }
	\label{text2motif}
\end{figure}

Since musical motifs are short in length, usually no more than one bar, they do not have obvious semantic information. The main attribution that dominates their listening perception is emotional information. Therefore our study focuses on generating musical motifs that have consistent emotional expression with the input text. Emotion of music has been shown to have a significant correlation with various musical features, such as pitch distribution and note density \cite{Meng2022}. By controlling these musical features, one can explicitly control the emotional expression of music. Consequently, the core problem to be solved in this work is how to generate these musical features from textual information. Previous research in music psychology has well-established the relationship between emotional valence and mode (major vs. minor), as well as musical structural features (e.g., tempo, fast or slow) in tonal music. Music scored in a major key is typically judged more positively, while music scored in a minor key is typically judged more negatively \cite{Heinlein1928}. Fast tempos tend to make music sound happier, while slow tempos tend to make music sound sadder \cite{Rigg1940}. Furthermore, faster tempos, higher note densities, and shorter note durations are associated with increased emotional arousal, while slower tempos, lower note densities, and longer note durations are associated with decreased arousal \cite{Thayer1990}. Consequently, Valence-Arousal (VA) emotion features are used as a link between textual information and musical features \cite{Forero2023Are, Livingstone2010}. Currently, the extraction of textual emotion information is a more mature technique \cite{Alswaidan2020, Deng2023}. In this work, a Textual Emotion Recognition (TER) pre-trained model is adopted to achieve the function of extracting VA values \cite{Mendes2023}. According to the above concept, TTMM is composed of two phases: the VA-based music feature extraction phase and the motif generation phase. The workflow of the TTMM is shown in Figure \ref{text2motif}.

In the first phase, mode, note density (ND), and note average duration (NAD) are chosen as the musical features, where mode refers to whether the music is in a major or minor key; ND is defined as the number of notes contained in one beat of time; and the NAD is the sum of all note durations divided by the number of notes in beats. To obtain these variable, the VA values is first predicted by the pre-trained TER model. Then the VA values are mapped to the musical features by an empirical approach. It first determines the range of each predicted value according to the music theory rules introduced in the above literature, and then obtains the target value by random sampling. Specifically, Valence determines the mode, and Arousal determines the ND and NAD. The detail is outlined in Algorithm \ref{VA_to_musical_feature}.

\begin{algorithm}[hbt]
	\caption{VA to musical feature mapping}
	\label{VA_to_musical_feature}
	\renewcommand{\algorithmicrequire}{ \textbf{Input:}} 
	\renewcommand{\algorithmicensure}{ \textbf{Output:}} 
	\begin{algorithmic}[1]
		\REQUIRE $Valence$, $Arousal$
		\ENSURE $mode$, $ND$, $NAD$
		
		\IF {$Valence \leq 5$} 
		\STATE $mode = Major$
		\ELSE
		\STATE $mode = Minor$
		\ENDIF
		\STATE $margin\_nd\gets[0, 3.5, 5, 8]$
		\STATE $margin\_nad\gets[0, 0.8, 1.2, 2]$
		\STATE $idx = CEIL(Arousal/3)$
		\STATE $ND$ = Randomly sampling between $margin\_nd[idx-1]$ and $margin\_nd[idx]$
		\STATE $NAD$ = Randomly sampling between $margin\_nad[4-idx]$ and $margin\_nad[3-idx]$
	\end{algorithmic}
\end{algorithm}

In the second phase, the Number of Notes (NoN) is first obtained by multiplying the ND value with the duration of one bar, and the NoN is adjusted to an integer value not less than 2 according to the compositional rules. Next, the list of tone differences is selected according to the given mode label, which is $ [0, 2, 4, 5, 7, 9, 11, 12]$ in the case of a major key or $[0, 2, 3, 5, 7, 8, 10, 12]$ in the case of a minor key. The number in the list indicates the semitone difference between the note within that mode and the reference note. For a given key, add the MIDI code of the base note to each element of the list to get the scale of the corresponding key. For example, if the mode label is "Major" and the key is "D", then add each element of the list to the MIDI code 62 of the base tone of D4 to get the scale list $[62, 64, 66, 67, 69, 71, 73, 74]$. Finally, the motif note sequence is determined by NoN times randomly choice from this list. On the other hand, the duration list is initialized by NAD, and then the length of the whole duration list is extended to one bar by randomly inserting beats (eighth note in our setup). Finally, the motif is obtained by combining the note sequence and the duration list.

\subsection{Melody Generation Module (MGM)}

\begin{figure}[hbt]
	\centering
	\centerline{\epsfig{figure=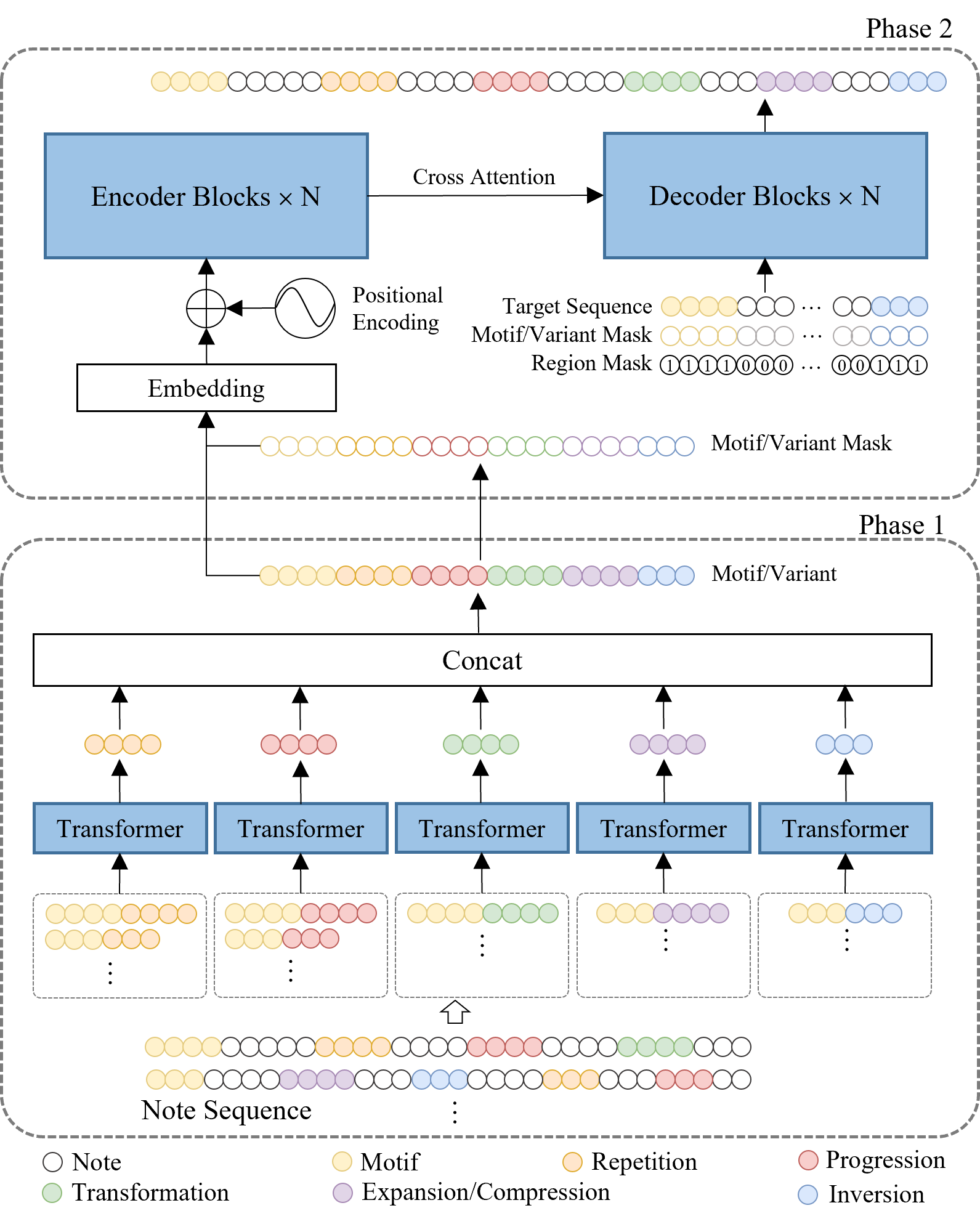,width=9cm}}
	\caption{Architecture of MGM. }
	\label{MGM}
\end{figure}

The MGM is designed to learn the motif-to-phrase transition for creating musical segments. It is also composed of two phases: the variants generation phase and the phrase generation phase. The architecture of the MGM is shown in Figure \ref{MGM}.

In the first phase, a multi-branch model is adopted to generate five types of variants, each branch specifically dedicated to learning one of the types. We use Theme Transformer\cite{shih_theme_2023} as backbone, which consists of an encoder and a decoder with gated parallel attention, using Region Mask to control the use of self-attention or cross-attention in the decoder. In the training stage, motif and variant fragments are extracted from the input token sequences to build motif-variant data pairs. Then these pairs are classified into five categories, each corresponding to a variant type. For each type, a Transformer takes motifs as encoder input and variants as decoder input and target. In the inference stage, this module deform the input motif to a specific variant. Then, all the variants are concatenated. Specifically: 
\begin{equation}
	V=\mathbf{Concat}(x,v_1,...,v_5),
\end{equation}
where $x$ indicates the input motif and  $v_j=\mathbf{TR_j}(x)$ represents Transformer for the $j$ type variant. This facilitates the clear and distinct learning of different variants, negating the potential for interference between different variants during the learning process. Furthermore, employing separate Transformers enhances the interpretability of the model, as it becomes easier to comprehend how each type of variation is applied. The resulting variants, combined with the original input motif, are subsequently transferred to the next phase.

\begin{figure}[hbt]
	\centering
	\centerline{\epsfig{figure=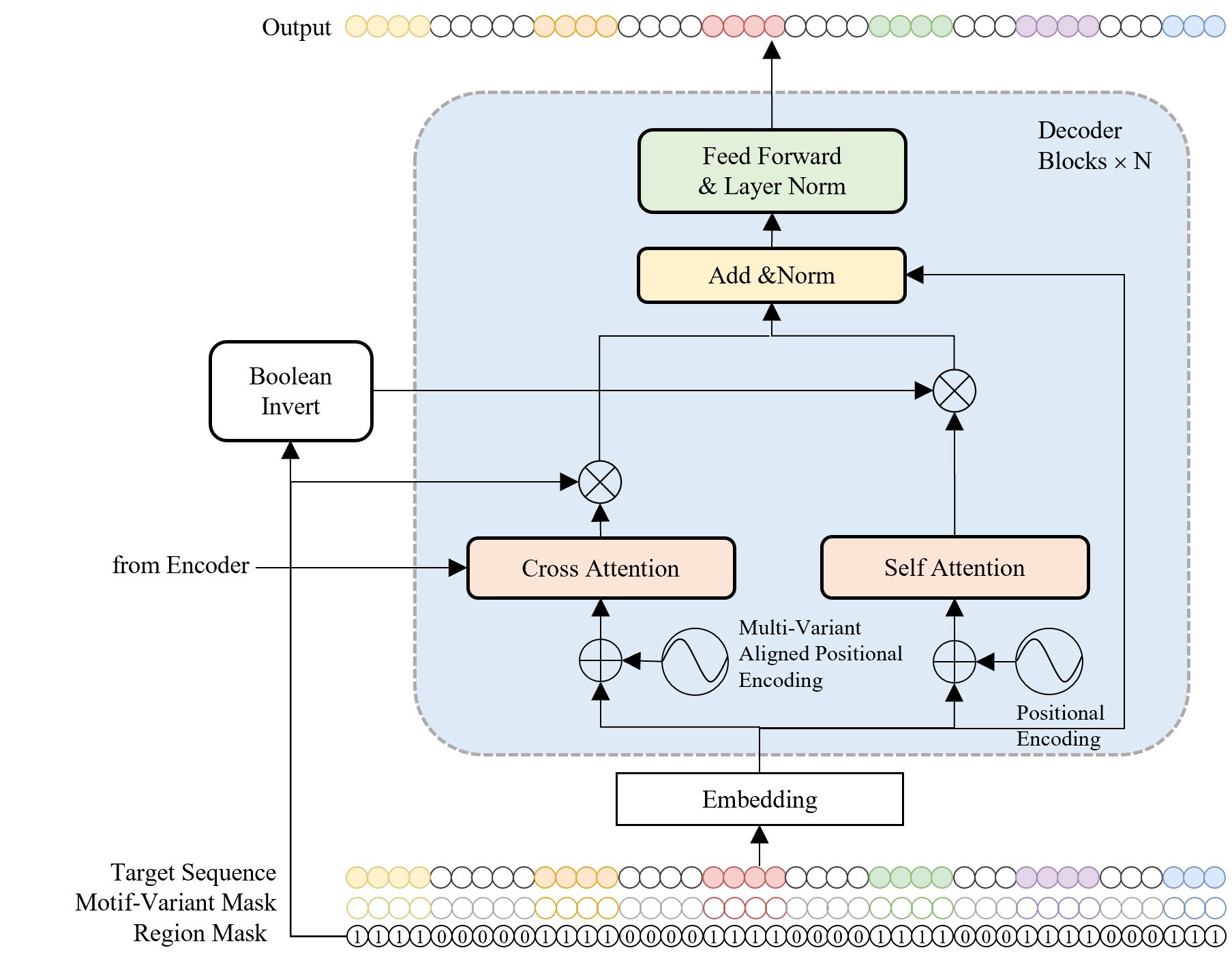,width=8.5cm}}
	\caption{The architecture of the decoder in MGM. }
	\label{fig:MGM_decoder}
\end{figure}

The second phase takes motif and variants as input and outputs musical phrase. Each motif label is processed as \textit{motif\_start} and \textit{motif\_end} tokens, while variant label as \textit{type}, \textit{motif\_start} and \textit{motif\_end} tokens. The \textit{type} and \textit{motif\_start} tokens are placed to the start of the motif/variant area, and the \textit{motif\_end} tokens to the end, dividing the target sequence into motif/variant areas and ordinary areas. The decoder only uses cross-attention when the current timestep \textit{t} is within a motif/variant area, otherwise only self-attention, as shown in Figure \ref{fig:MGM_decoder}. To distinguish between motif and various types of variants, we introduce a Motif/Variant Mask to provide unique positional encoding information for different variants. Furthermore, a Multi-Variant Aligned Positional Encoding (MVAPE) is proposed to synchronize the positional encoding of motif/variant area in the decoder with that required for cross-attention. The decoder process is given as:
\begin{equation}
	h_t^c=\mathbf{Emb}(y_t)+\mathbf{MVAPE}(y_t,m_t),
\end{equation}
\begin{equation}
	h_t^s=\mathbf{Emb}(y_t)+\mathbf{PE}(y_t),
\end{equation}
\begin{equation}
	h_t=r_t*\mathbf{CrossAttn}(h_t^c,h_{enc})+(1-r_t)*\mathbf{SelfAttn}(h_t^s),
\end{equation}
where $m_t$ and $r_t$ indicate Motif/Variant Mask and Region Mask at timestep $t$, respectively. $h_{enc}$ represents the encoder output.

\subsubsection{Motif/Variant Mask}
The Motif/Variant Mask $\textbf{m}=\{m_1,m_2…m_i\}$ aims to distinguish between motif and different variant types. It assigns a distinct mask to each motif/variant, providing positional information for the MVAPE. Specifically, for the $i$-th token $t_i$ in the sequence, we define:
\begin{equation}
	m_i=
	\begin{cases}
		j*2l_m+k,&t_i \in mv\\
		0,&t_i \notin mv
	\end{cases}
\end{equation} 
where  $mv$ indicates the motif/variant area, and $j\in\{0,1,\ldots,5\}$ represents the area type: 0 for motif and 1–5 for variants.  $l_m$ signifies the length of the motif's token sequence, and $k$ is the index of the token in the motif/variant sequence, satisfying $k<2l_m$. This ensures the uniqueness of each Motif/Variant Mask, preventing any duplication in the mask sequence.

\subsubsection{Multi-Variant Aligned Positional Encoding (MVAPE)}

The Multi-Variant Aligned Positional Encoding (MVAPE) seeks to align the positional encoding ($PE$) of the decoder input motif/variant area with the $PE$ of the input sequence for cross-attention. For instance, when a \textit{motif\_start} token is generated for a Progression (variant type $j=2$) at timestep $t$, the decoder may encounter difficulties in cross-attending to tokens within the input sequence's Progression area when predicting timestep $t+1$. This challenge stems from the substantial positional discrepancy in the standard positional encoding. Although Theme Transformer\cite{shih_theme_2023} achieves a "clock reset" by providing Region Mask to align $PE$s, it fails to differentiate among different variant types. To address this limitation, we propose the MVAPE. Specifically, when a \textit{motif\_start} token of a type $j$ variant is generated at timestep $t_j$, and $\rho$ represents the index of the $i$-th token in the variant area, $PE_{j}(\rho)$ is employed as the positional encoding for timestep $i$ to calculate cross-attention when predicting subsequent tokens, continuing until the decoder generates a \textit{motif\_end} token. Specifically:
\begin{equation}
	\mathbf{PE}(i)=\mathbf{PE}(t_j+\rho)=\mathbf{PE}_j(\rho)
\end{equation}
In this context, $PE_j$ represents the positional encoding of the type $j$ variant in the input sequence, which is extractable through the Motif/Variant Mask. Consequently, the MVAPE not only maintains the advantages of computing cross-attention but also effectively differentiates among various variant types.

\section{Experiments}

\subsection{Experiment Setup}

\begin{figure*}[hbt]
	\centering
	\centerline{\epsfig{figure=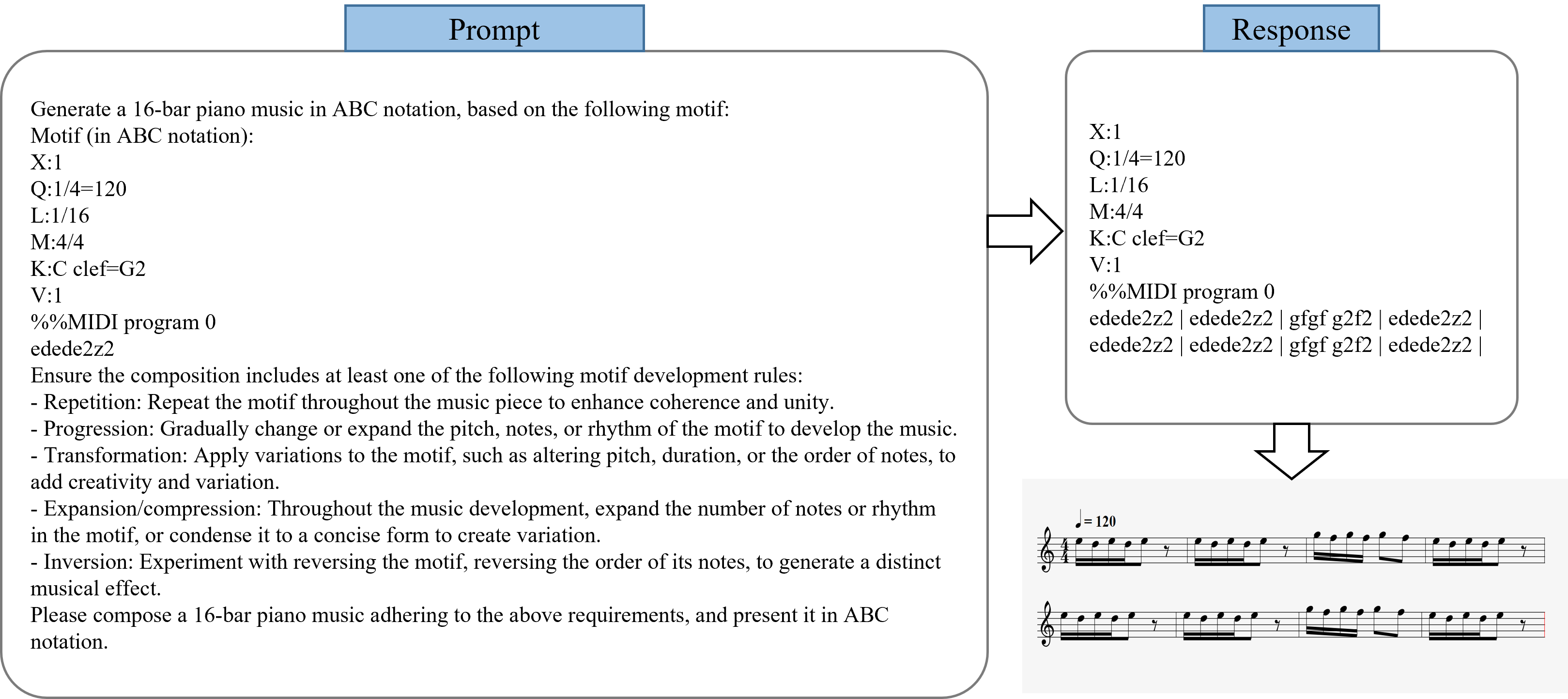,width=15cm}}
	\caption{ChatGPT-4 prompt for motif-to-music task. }
	\label{fig:chatgpt4prompt}
\end{figure*}

\begin{figure*}[hbt]
	\centering
	\centerline{\epsfig{figure=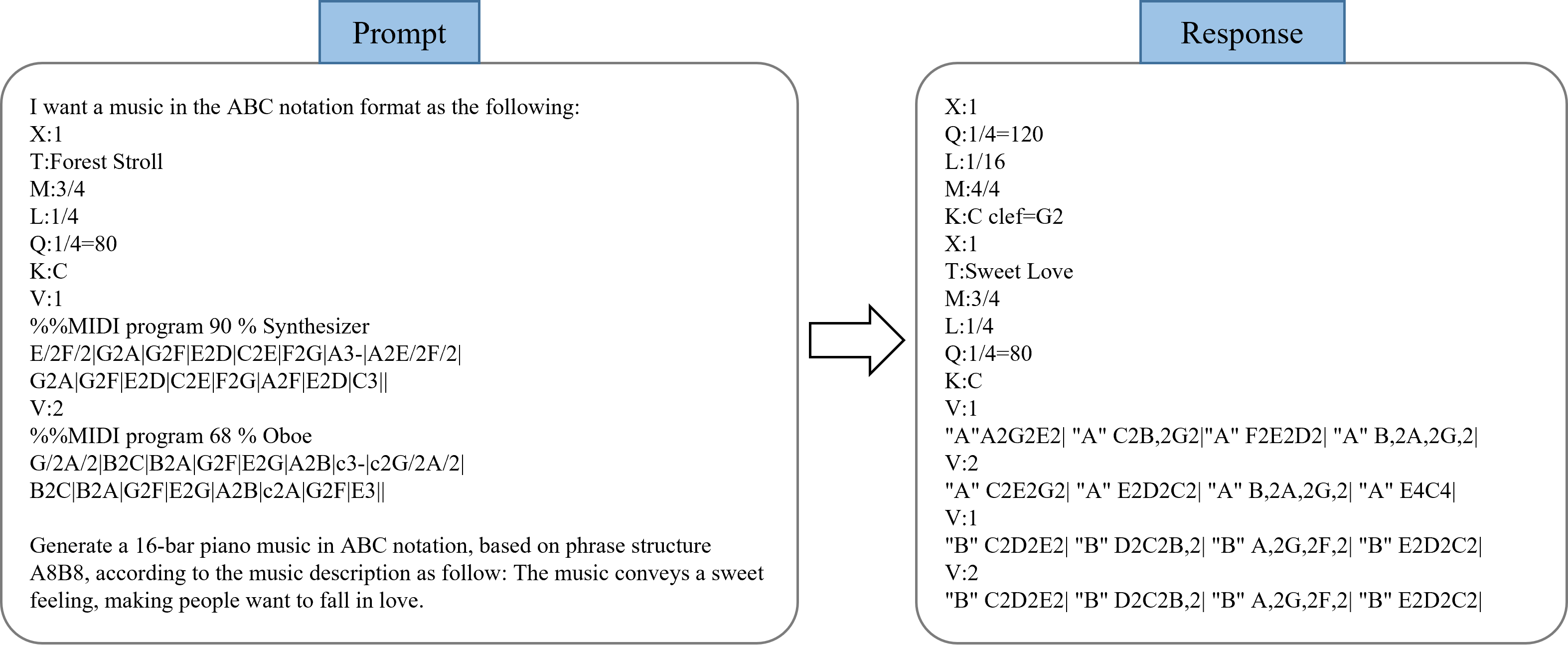,width=14cm}}
	\caption{ChatGPT-4 prompt for text-to-music task. }
	\label{fig:chatgpt4prompt2}
\end{figure*}

The experiments in this work are divided into three parts, which are dataset-specific experiments, comparison experiments and ablation experiments. In the first part, since there is no existing work that uses motif development strategy to generate music, we train Theme Transformer \cite{shih_theme_2023} on the POP909\_M dataset and perform comparison experiments with the MGM module proposed in this paper. ChatGPT-4 \cite{openai2024gpt4} can also be used to synthesize music clips from motifs, so we also adopt it for comparison in this experiment.To achieve this, the input motifs are converted to ABC notation \cite{walshaw_abcnotation_1995} and the output is converted to in MIDI format. Prompt used to instruct ChatGPT-4 is shown in Figure \ref{fig:chatgpt4prompt}.
In the second part, both objective and subjective assessments were adopted for comparison. Because there are not many open-source systems for generating symbolic music from text, within the scope of our endeavor, we use the following models as comparative models: Compound Word Transformer (CWT) \cite{hsiao_compound_2021},  Textune \cite{Wu2023}, MuseCoco \cite{Lu2023}, and ChatGPT-4 \cite{openai2024gpt4}. The CWT is not a text-to-music generative model. In order to meet our requirements, we add a BERT \cite{Devlin2019BERTPO} model as the encoder to form an Encoder-Decoder structure and train it on the POP909\_M dataset. The BERT model uses the Chinese\_L-12\_H-768\_A-12 checkpoint avaliable on Google website$\footnote{https://github.com/google-research/bert}$ and frozen in training process. Since Textune does not give open source code and MuseCoco does not give a suitable dataset for training, we just download the given synthetic clips from their respective official webpages$\footnote{https://huggingface.co/spaces/sander-wood/text-to-music}$$\footnote{https://ai-muzic.github.io/musecoco}$ for comparison. To reach the text-to-music synthesis task, ChatGPT-4's prompt is shown in Figure \ref{fig:chatgpt4prompt2}.
In the third part, in order to verify the effectiveness of the MVAPE proposed in this paper, we do ablation experiments for it.

All the MIDI files were converted to REMI tokens\cite{huang_pop_2020}. Motif and variant labels were processed into tokens and subsequently incorporated into the REMI sequence. Notes were quantized to sixteenth note precision. The dataset was randomly partitioned into training, validation, and test sets with a ratio of 8.5:1:0.5, respectively.

All Transformer models utilized a 6-layer encoder and 6-layer decoder, featuring 8 heads for multi-head attention. The maximum sequence length is 1024, and embedding size is 256. The dimension of the feed-forward layer was set to 2048. The model was optimized using the Adam optimizer ($\beta_1=0.9, \beta_2=0.99$) to minimize the training negative log-likelihood, employing an initial learning rate of 0.0002 and a batch size of 4. The model was trained for 1000 epochs using an NVIDIA GeForce RTX 2080 Ti GPU.

\subsection{Objective Measures}

Because symbolic music generation lacks standardized objective evaluation metrics due to the intrinsically subjective nature of music and the difficulty in quantifying its qualities. In order to explore the learning capability on the motif-to-phrase pattern, we assess whether the model generates variants correctly, and whether the distribution of variant type is reasonable. Two novel objective evaluation metrics in the objective experiment are proposed: \textit{Variant Proportion} ($\mathbf{VP_i}$) calculates the proportion of the $i$-th type of variant and \textit{Variant Distance} ($\mathbf{VD}$) calculates the average length (in beats) between motifs or variants. Specifically: 
\begin{equation}
	\mathbf{VP_i}=(\sum_{j=1}^{n_d}n_i )/N,
\end{equation}
where $n_d$ denotes the number of samples, $n_i$ represents the \textit{$i$}-th type of variant in the current dataset.
\begin{equation}
	\mathbf{VD}=(\sum_{j=1}^{n_d}\sum_{k=2}^{n_v}t_k^s-t_{k-1}^s)/N,
\end{equation}
where $n_v$ denotes the number of variants in the current sample, and $t_k^s$ represents the start time of the $k$-th variant. The proximity of each score to that of the dataset indicates a superior effect.

\subsection{Dataset Evaluation}

To assess the impact of our proposed dataset, we include the original Theme Transformer trained on POP909 for comparison. All samples are presented in monophonic form. From the results in Table \ref{Dataset_Objective_Evaluation}, it can be observed that without motif and variant labels, Theme Transformer trained on POP909 dataset fail to acquire the understanding of diverse variants, only show ability to generate Repetitions variant. In contrast, models trained on POP909\_M dataset has shown the ability to capture information of different kinds of variant. ChatGPT-4 follows a strict instruction to generate "repetition" and "progression" all the time without additionally connection, resulting in low score on $\mathbf{VD}$. Overall, MeloTrans performs a closer distribution of variant types in generated samples compared to the real data. These experimental results illustrate that the POP909\_M dataset can effectively improve the structure of synthesised music. The structural diversity of the generated music can be further enhanced by adopting suitable model structures. The LLM on massive data do not effectively obtain these structural information, which in a way validates the conjecture at the beginning of this paper.
\begin{table}[hbt]
	\centering
	\caption{Dataset evaluation results.}
	\label{Dataset_Objective_Evaluation}
	\begin{tabularx} {0.5\textwidth}{X m{0.5cm} m{0.5cm} m{0.5cm} m{0.5cm} m{0.5cm} m{0.5cm}}
		\hline
		\textbf{Model} & $\mathbf{VP_1}$ & $\mathbf{VP_2}$ & $\mathbf{VP_3}$ & $\mathbf{VP_4}$ & $\mathbf{VP_5}$ & $\mathbf{VD}$ \\
		\hline
		POP909\_M & 0.22 & 0.12 & 0.11 & 0.51 & 0.04 & 7.73 \\
		\hline
		Theme Transformer (POP909) & 0.86& 0& 0& 0.14& 0&9.86\\
		Theme Transformer (POP909\_M) & 0.17& \textbf{0.13}& 0.02& 0.68& 0&\textbf{8.90}\\
		ChatGPT-4 & 0.54& 0.28& 0.02& 0.16& 0&4.25\\
		MGM (ours) & \textbf{0.19} & 0.14 & \textbf{0.06} & \textbf{0.60} & \textbf{0.01} & 9.07 \\
		\hline
	\end{tabularx}
\end{table}

\subsection{Subjective Evaluation}

Twenty participants are invited to do the subject evaluation. Ten of them are professionals in music while others are not. The test set consists of 30 music clips, 25 of which are composed of five pieces of music generated by each of the five test models, and five randomly selected pieces of real music from the POP\_909M dataset for comparison. The order of the samples is randomized for each listener. The scoring is based on a 5-point Likert scale \cite{likert_technique_1932}, the higher the better.
We define the following metrics: 
\begin{enumerate}[itemsep=0pt,topsep=3pt,parsep=0pt, partopsep=0pt]
	\item Musicality (\textbf{M}): The degree to which the generated music is similar to human composition in listening;
	\item Structure (\textbf{S}): whether the generated music has musical structure, e.g. recognisable development of motifs-to-variants and reasonable arrangement of phrases.
	\item Semantic Matching Degree (\textbf{SMD}): How well the generated music matches the expressiveness described by the input text;
	\item Overall Evaluation(\textbf{OE}): the evaluation of the overall situation of the generated music;
\end{enumerate}

\begin{table}[hbt]
	\centering
	\caption{Subjective evaluation results.}
	\label{Subjective_Evaluation}
	\begin{tabular}{lllll}
		\hline
		\textbf{Model (year)} & \textbf{M} & \textbf{S} & \textbf{SMD} & \textbf{OE} \\
		\hline
		POP909\_M & 4.26 & 4.12 & 3.86 & 4.08 \\
		\hline
		CWT (2021) & 3.34 & 2.98 & 3.02 & 3.25 \\
		Textune (2023) & 2.32 & 2.08 & 2.15 & 2.21 \\
		MuseCoco (2023) & \textbf{4.06} & 3.67 & 3.26 & 3.65 \\
		ChatGPT-4 (2023) & 2.65 & 3.07 & 3.10 & 3.05 \\
		\textbf{ours} & 3.72 & \textbf{3.88} & \textbf{3.39} & \textbf{3.69} \\
		\hline
	\end{tabular}
\end{table}

The Table \ref{Subjective_Evaluation} shows that our model gets the highest scores on all metrics except \textbf{M}, which is also close to the highest scoring MuseCoco. Because MuseCoco was trained on nearly 1 million text-music pairs, which were obtained by mixing multiple datasets, and thus MuseCoco has a great advantage in musicality, but our model also achieves better scores with much smaller data sets. On the other hand, MuseCoco's text is obtained by combining labels and templates from different datasets through ChatGPT. The labels mainly focus on low-level features (such as number of bars, BPM, Musical Instruments, etc.), so the control of high-level semantic features (such as sentiment, genre) is not precise. The generation controllability of MuseCoco will decrease somewhat. The model in this paper, on the other hand, predicts low-level musical features from high-level semantic information, and considers the control of different levels of features on the generation process, thus obtaining a high \textbf{SMD} score. In terms of \textbf{S}, MuseCoco still loses to our model due to the fact that it does not consider structural modelling, although it learns some structure from the massive data. Although ChatGPT-4 can be instructed to generate music by designing prompt, it cannot obtain enough detailed information because user descriptions usually use short text, and thus the generated music tends to sound monotonous and lacks variation in rhythm patterns, resulting in low scores. Textune can only generate monophonic melody in many cases, and tends to repeat a few notes or use the sequence from do to si, which is neither sufficient to express the semantics of the textual description nor form a reasonable musical structure, and thus scores poorly in all categories; Due to the limited amount of training data and no additional feature extraction measures, Compound Word Transformer also has low scores, which is consistent with the results shown in objective evaluation experiments. In terms of \textbf{OE} scores, MuseCoco is still inferior to our model. This shows that combining human composition knowledge with the transformer model can effectively improve the performance of automatic composition, much more effectively than simply training with massive amounts of data.

\subsection{Objective Evaluation}

Since the metrics used in the objective experiments need to be compared with the data measured from the dataset, the comparison model must be able to train with POP909\_M dataset, which can be used to observe whether the test model can learn the distribution of real data. If the model is trained with other datasets, it will not be able to fairly compare the learning ability of the models due to the difference in the distribution of the datasets. Textune and MuseCoco do not have such a condition, so in this experiment, we only compare with CWT. To show this difference, we also include ChatGPT-4 for comparison.

\begin{table}[hbt]
	\centering
	\caption{ Objective evaluation results.}
	\label{Objective_evaluation}
	\begin{tabular}{l l l l l l l}
		\hline
		\textbf{Model} & $\mathbf{VP_1}$ & $\mathbf{VP_2}$ & $\mathbf{VP_3}$ & $\mathbf{VP_4}$ & $\mathbf{VP_5}$ & $\mathbf{VD}$ \\
		\hline
		POP909\_M & 0.22 & 0.12 & 0.11 & 0.51 & 0.04 & 7.73 \\
		\hline
		CWT & 0.19 & 0.07 & 0.34 & 0.39 & 0.01 & \textbf{7.39} \\
		ChatGPT-4 & 0.28 & 0.21 & 0.37 & 0.14 & 0 & 4.16 \\
		\textbf{ours} & \textbf{0.24} & \textbf{0.10} & \textbf{0.18} & \textbf{0.47} & \textbf{0.01} & 8.65 \\
		\hline
	\end{tabular}
\end{table}
The results of the objective experiment is shown in Table \ref{Objective_evaluation}. It can be seen that the proposed model is closer to the real data in terms of the distribution of the variants. Our model learns the probability weights of different variant branches in the training process, which achieves a more reasonable distribution of the number of variants. However, our model is weaker than CWT on the $\mathbf{VD}$ data, which represents the variant distance. This may be due to the fact that our model generates music at a fixed 16 bars, whereas CWT predicts the \textit{end} token to decide when to stop generation. This makes the music generated by CWT potentially more compact and therefore shorter in the distance between motives/variants. But for normal music, this value is all reasonable around 8. That is, the average distance between variants is about two bars, which is in line with the musical experience and cognition that the length of a phrase is generally about two bars.

ChatGPT-4's performance on the text-to-music task differs from that on the motif-to-music task. Except for the inability to generate the "inverse" variant, the distribution of the other variants is no longer imbalanced as in Table \ref{Dataset_Objective_Evaluation}. However, its variant distribution still has a large gap from the true value. In addition, it still lacks additional connecting phrases, so the $\mathbf{VD}$ value is far from the real data which out of the normal range. Because ChatGPT-4 is trained on a different dataset and generated in a different way, the music generated by ChatGPT-4 is structurally different from our model. This makes it unsuitable for the comparison of this subsection. This also suggests that even ChatGPT-4, trained on a large amount of data, is difficult to learn some of the motif development patterns in music. In this experiment, there is a vacancy in the "inverse" mode.  It also shows the value of our development motivation strategy.

\subsection{Ablation Study}
To analyze the impact of MVAPE, we assess the performance of the MGM using vanilla positional encoding instead and without Motif/Variant Mask. The results in Tables \ref{Ablation_Study} shows that both methods degrade the performance on $\mathbf{VD}$ and the distribution of variant types. This indicates that the proposed MVAPE contributes to learning the appropriate timing for placing a suitable variant. 
\begin{table}[hbt]
	\centering
	\caption{ Ablation study results.}
	\label{Ablation_Study}
	\begin{tabular}{l l l l l l l}
		\hline
		\textbf{Model} & $\mathbf{VP_1}$ & $\mathbf{VP_2}$ & $\mathbf{VP_3}$ & $\mathbf{VP_4}$ & $\mathbf{VP_5}$ & $\mathbf{VD}$ \\
		\hline
		POP909\_M & 0.22 & 0.12 & 0.11 & 0.51 & 0.04 & 7.73 \\
		\hline
		Vanilla PE & 0.25 & 0.02 & 0.01 & 0.72 & 0 & 6.03 \\
		w/o mask& 0.17 & 0.09 & 0.04 & 0.70 & 0 & 5.24 \\
		MGM & \textbf{0.19} & \textbf{0.14} & \textbf{0.06} & \textbf{0.60} & \textbf{0.01} & \textbf{9.07} \\
		\hline
	\end{tabular}
\end{table}

It can be observed that the model without Motif/Variant Mask also generates a generally reasonable distribution of variant types. Therefore, we further analyze the correctness by comparing the generated variant \textit{type} token with the real type and calculating the proportion of matched ones. 
\begin{figure}[htb]
	\centering
	\subfloat[]{
		\includegraphics[width=3.7cm,height=3cm]{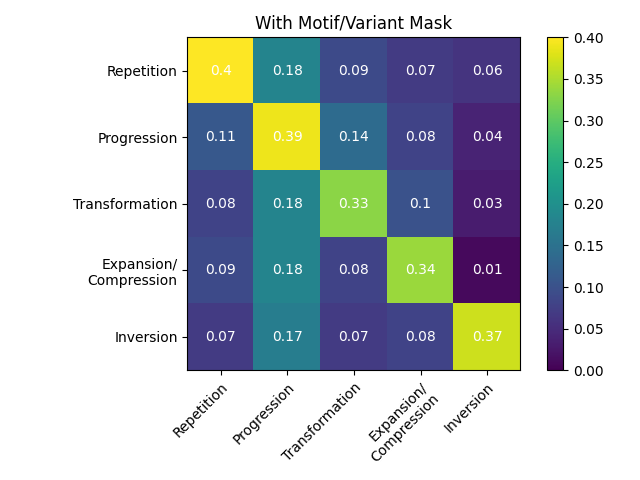}
	}
	\hfil
	\subfloat[]{
		\includegraphics[width=3.7cm,height=3cm]{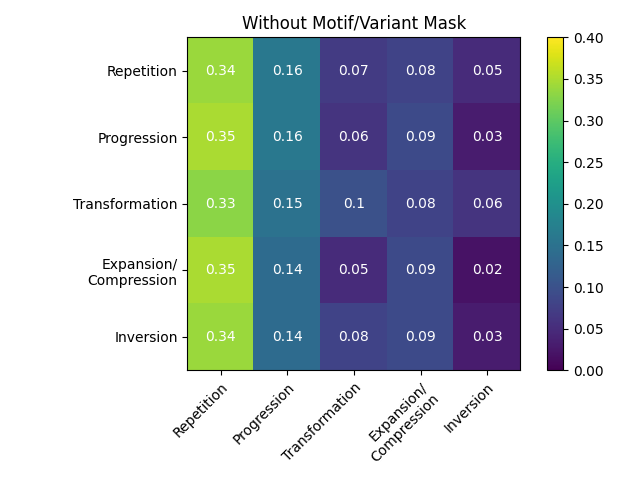}
	}
	
	\caption{Variant type distribution comparison of \textbf{(a)} with Motif/Variant Mask and \textbf{(b)} without Motif/Variant Mask.}
	\label{fig:Variant_type_distribution}
\end{figure}
Results of the model with/without Motif/Variant Mask are shown in Figure \ref{fig:Variant_type_distribution}. The row represents the type of generated variant marked by \textit{type} token, and the column represents the real type. For instance, 0.11 in the first column of the second row indicates that 11\% of the the marked "progression" in the generated samples are actually "repetition". It can be observed that the proposed Motif/Variant Mask has a beneficial impact on distinguishing different types of variants, achieving certain consistency between marked type and the real type of output variants. In contrast, the model without Motif/Variant Mask generates significant proportion of "repetition" regardless of the generated variant type token. Moreover, a certain proportion of the variants does not belong to any variant types specified in this paper, increasing the diversity of the generated music.

\subsection{Comparison of generation results}

\begin{figure*}[htb]
	\centering
	\subfloat[]{
		\includegraphics[width=12cm,height=3.5cm]{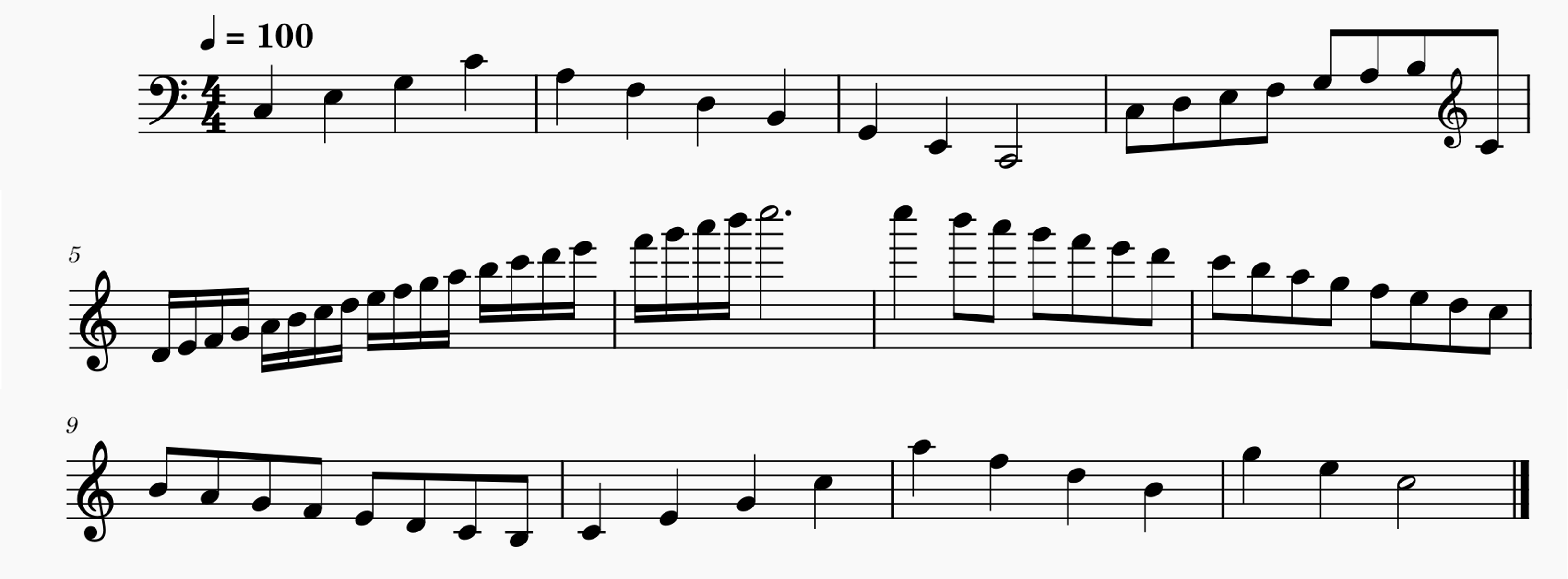}
	}
	\hfil
	\subfloat[]{
		\includegraphics[width=12cm,height=3.5cm]{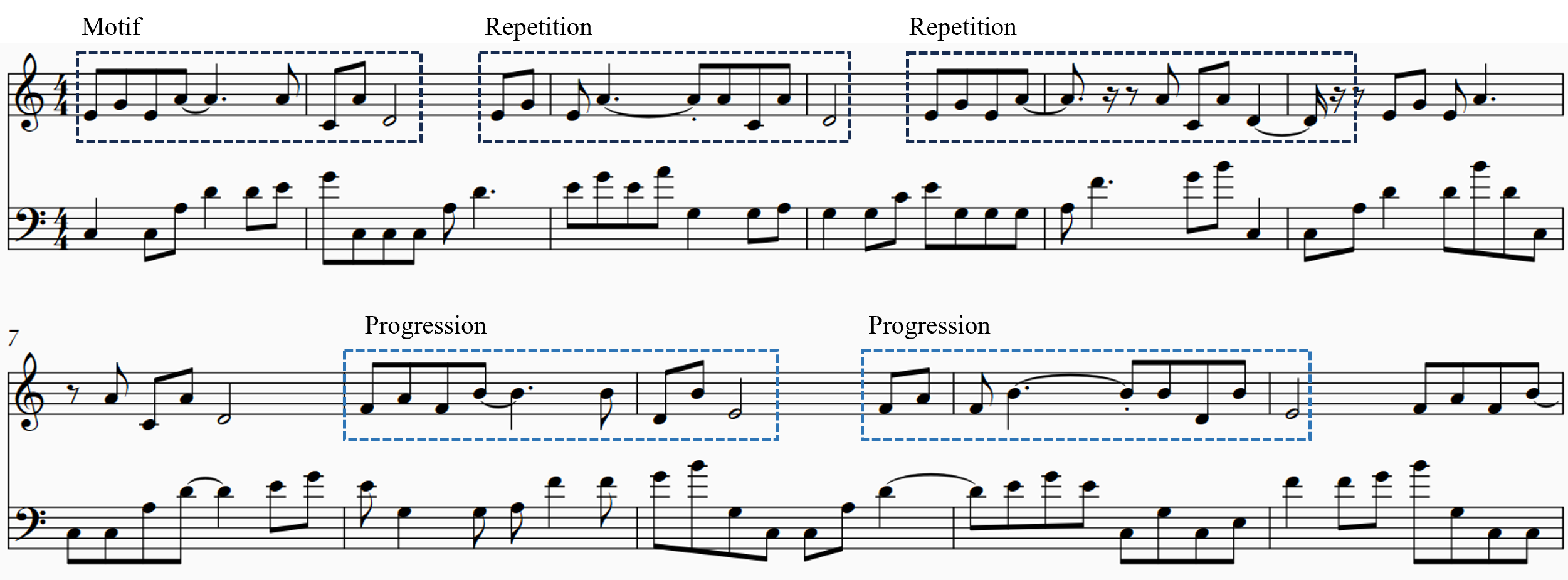}
	}
	\hfil
	\subfloat[]{
		\includegraphics[width=12cm,height=3.5cm]{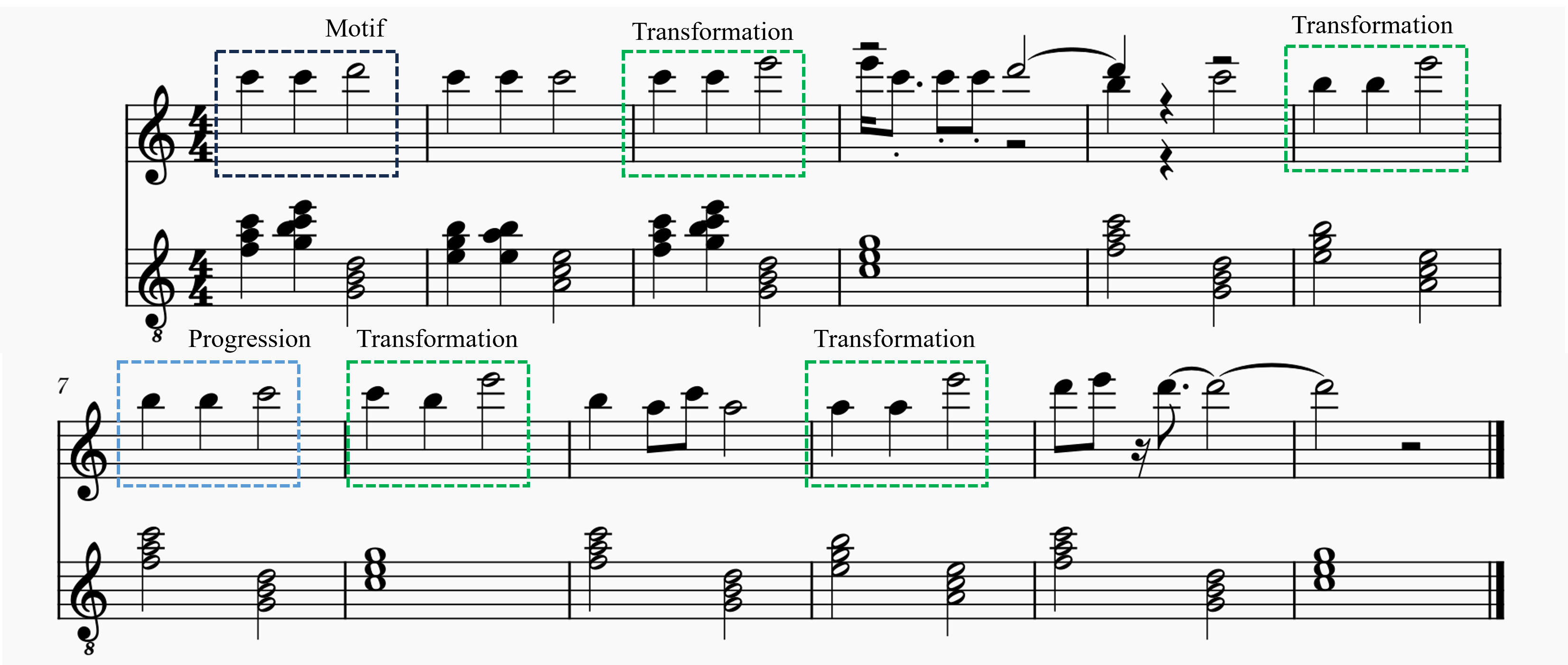}
	}
	\hfil
	\subfloat[]{
		\includegraphics[width=12cm,height=3.5cm]{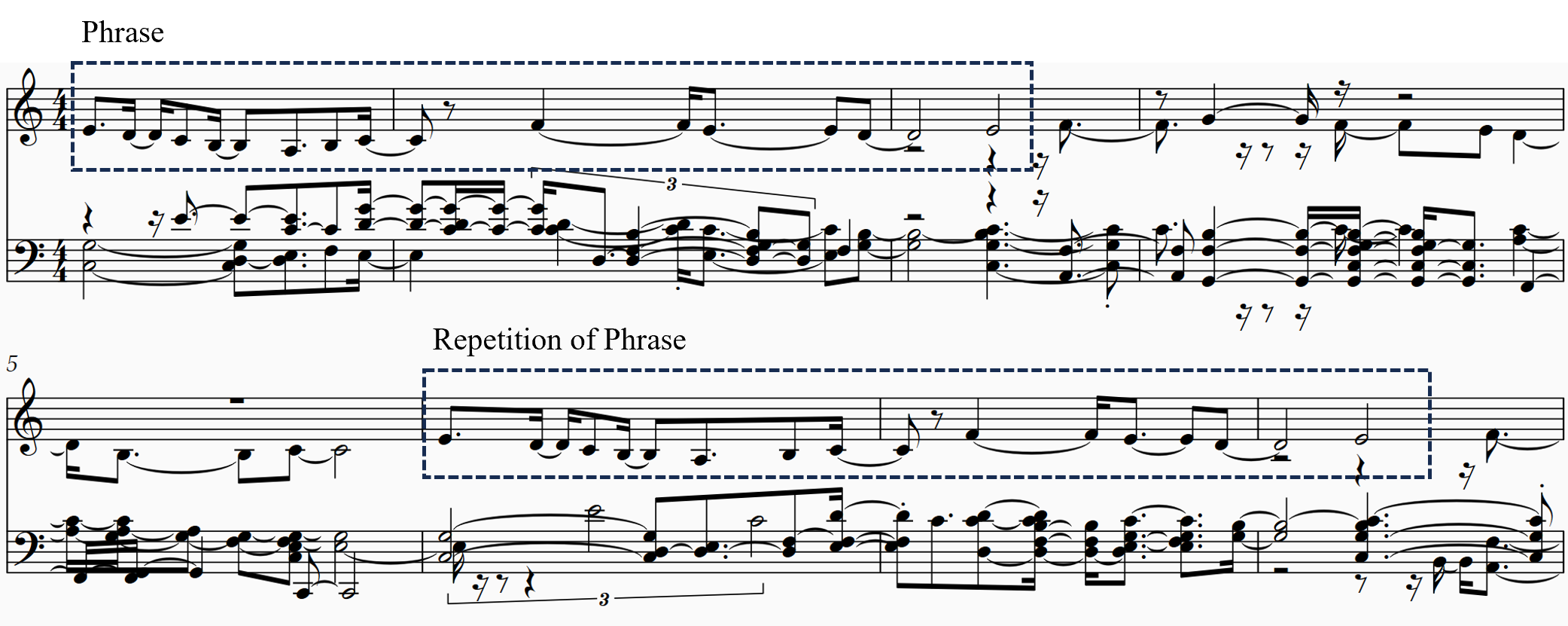}
	}
	\hfil
	\subfloat[]{
		\includegraphics[width=12cm,height=4cm]{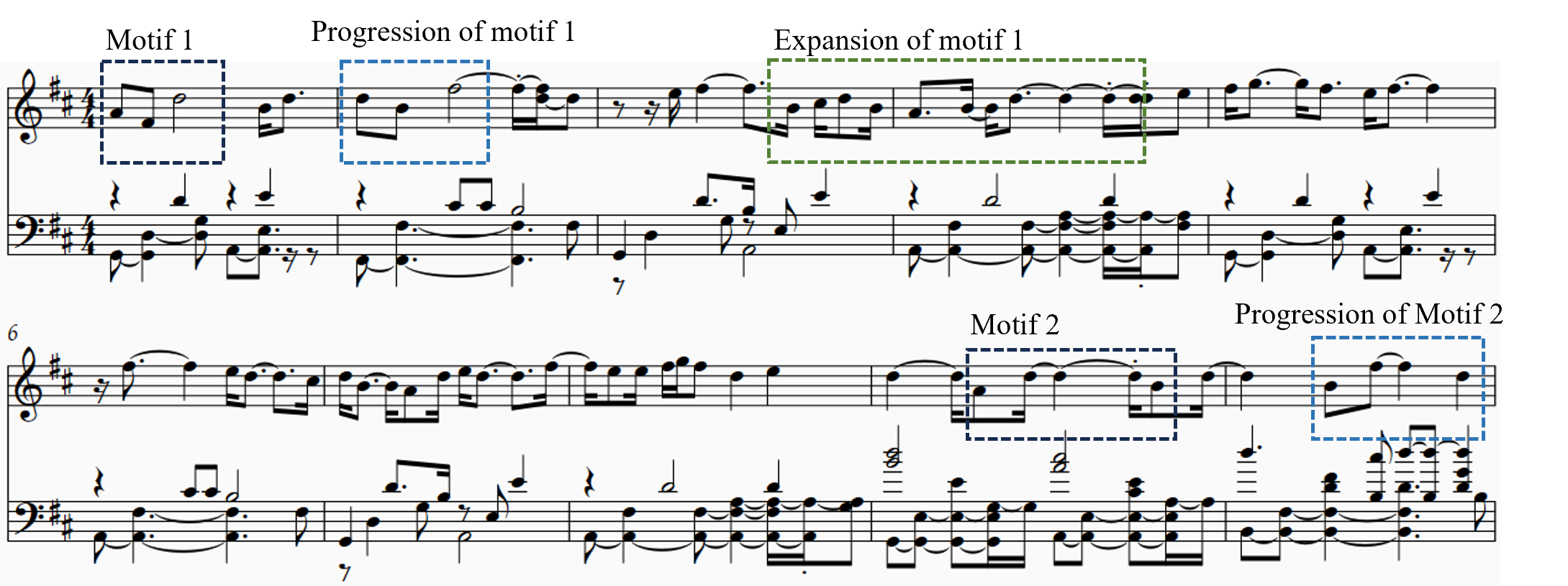}
	}	
	\caption{Generation results comparison of \textbf{(a)} Textune, \textbf{(b)} ChatGPT-4, \textbf{(c)} Compound Word Transformer, \textbf{(d)} MuseCoco, \textbf{(e)} our model.}
	\label{fig:stave_comparison}
\end{figure*}

In order to show the performance of music generated by different models more intuitively, we select samples of each model and demonstrated in Figures \ref{fig:stave_comparison}(a)-(e), which are the music generated by Textune, ChatGPT-4, Compound Word Transformer, MuseCoco, and the algorithm in this paper, respectively. The input text is described as "This is a romantic and sweet song". It can be seen that the music generated by Textune is monophonic, with no obvious phrase structure, and the arrangement of notes is very monotonous. The music generated by ChatGPT-4 has obvious "repetition" and "transformation" variants, but there is no suitable phrase connection between them; Compound Word Transformer can learn some motif development rules from the data, and can find "transformation" and "progression" patterns in the melody. However, the melody is suffered from relatively monotonous. the music generated by MuseCoco has a “repetition” pattern, but the motif within the phrase is not obvious. The “repetition” pattern is based on a full phrase but not motif, which is not a clear pattern. The music generated by our model is relatively well structured. There are multiple motifs in the melody, and the motifs are developed in a variety of ways. The connections between these variants are also well tuneful. This proves that the our model can generate music with clearer structured and richer melody.

\section{Conclusion}
In this paper, we attempt to combine deep learning model with human composition rules. We consider motif development as the key of human composition and propose POP909\_M dataset with motif, variant and text description. We introduce MeloTrans, a music composition model based on motif development. Experimental results demonstrate that our model successfully follows motif development rules while achieving greater musical richness compared to the baselines.

Future extensions of this work could involve following additional steps of human composition (e.g., combining multiple segments and incorporating formal structure to generate complete music), thereby enabling more comprehensive and longer compositions. This will contribute to the continuous evolution of symbolic music generation and its applications. 

\bibliographystyle{IEEEtran}

\bibliography{ref2}

\begin{thebibliography}{10}
\providecommand{\url}[1]{#1}
\csname url@samestyle\endcsname
\providecommand{\newblock}{\relax}
\providecommand{\bibinfo}[2]{#2}
\providecommand{\BIBentrySTDinterwordspacing}{\spaceskip=0pt\relax}
\providecommand{\BIBentryALTinterwordstretchfactor}{4}
\providecommand{\BIBentryALTinterwordspacing}{\spaceskip=\fontdimen2\font plus
\BIBentryALTinterwordstretchfactor\fontdimen3\font minus
  \fontdimen4\font\relax}
\providecommand{\BIBforeignlanguage}[2]{{%
\expandafter\ifx\csname l@#1\endcsname\relax
\typeout{** WARNING: IEEEtran.bst: No hyphenation pattern has been}%
\typeout{** loaded for the language `#1'. Using the pattern for}%
\typeout{** the default language instead.}%
\else
\language=\csname l@#1\endcsname
\fi
#2}}
\providecommand{\BIBdecl}{\relax}
\BIBdecl

\bibitem{roberts_hierarchical_2018}
A.~Roberts, J.~Engel, C.~Raffel, C.~Hawthorne, and D.~Eck, ``A {Hierarchical}
  {Latent} {Vector} {Model} for {Learning} {Long}-{Term} {Structure} in
  {Music},'' in \emph{Proceedings of the 35th {ICML}}, Jul. 2018, pp.
  4364--4373.

\bibitem{huang_music_2018}
C.-Z.~A. Huang, A.~Vaswani, J.~Uszkoreit, N.~M. Shazeer, I.~Simon,
  C.~Hawthorne, A.~M. Dai, M.~D. Hoffman, M.~Dinculescu, and D.~Eck, ``Music
  transformer: Generating music with long-term structure,'' in
  \emph{International Conference on Learning Representations}, 2018.

\bibitem{hsiao_compound_2021}
W.~Y. Hsiao, J.~Y. Liu, Y.~C. Yeh, and Y.~H. Yang,
  ``\BIBforeignlanguage{en}{Compound {Word} {Transformer}: {Learning} to
  {Compose} {Full}-{Song} {Music} over {Dynamic} {Directed} {Hypergraphs}},''
  in \emph{\BIBforeignlanguage{en}{Proceedings of the AAAI Conference on
  Artificial Intelligence}}, May 2021, pp. 178--186.

\bibitem{zhao_accomontage_2021}
J.~Zhao and G.~G. Xia, ``Accomontage: Accompaniment arrangement via phrase
  selection and style transfer,'' in \emph{International Societyfor Music
  Information Retrieval Conference}, 2021.

\bibitem{shih_theme_2023}
Y.~J. Shih, S.~L. Wu, F.~Zalkow, M.~Müller, and Y.~H. Yang, ``Theme
  {Transformer}: {Symbolic} {Music} {Generation} {With} {Theme}-{Conditioned}
  {Transformer},'' \emph{IEEE Transactions on Multimedia}, vol.~25, pp.
  3495--3508, 2023.

\bibitem{yi_accomontage2_2022}
L.~Yi, H.~Hu, J.~Zhao, and G.~G. Xia, ``Accomontage2: A complete harmonization
  and accompaniment arrangement system,'' in \emph{International Society for
  Music Information Retrieval Conference}, 2022.

\bibitem{wu_popmnet_2020}
J.~Wu, X.~Liu, X.~Hu, and J.~Zhu, ``{PopMNet}: {Generating} structured pop
  music melodies using neural networks,'' \emph{Artificial Intelligence}, vol.
  286, p. 103303, Sep. 2020.

\bibitem{zou_melons_2022}
Y.~Zou, P.~Zou, Y.~Zhao, K.~Zhang, R.~Zhang, and X.~Wang, ``Melons:
  {Generating} {Melody} {With} {Long}-{Term} {Structure} {Using} {Transformers}
  {And} {Structure} {Graph},'' in \emph{{ICASSP} 2022}, May 2022, pp. 191--195.

\bibitem{wu_power_2023}
G.~Wu, S.~Liu, and X.~Fan, ``The {Power} of {Fragmentation}: {A} {Hierarchical}
  {Transformer} {Model} for {Structural} {Segmentation} in {Symbolic} {Music}
  {Generation},'' \emph{IEEE/ACM Transactions on Audio, Speech, and Language
  Processing}, vol.~31, pp. 1409--1420, 2023.

\bibitem{lu_meloform_2022}
P.~Lu, X.~Tan, B.~Yu, T.~Qin, S.~Zhao, and T.~Y. Liu, ``{MeloForm}:
  {Generating} {Melody} with {Musical} {Form} based on {Expert} {Systems} and
  {Neural} {Networks},'' in \emph{23rd Int. Society for Music Information
  Re-trieval Conf. (ISMIR)}, Bengaluru, India, Aug. 2022, arXiv:2208.14345.

\bibitem{kachulis_songwriters_2003}
K.~J., \emph{\BIBforeignlanguage{en}{The {Songwriter}'s {Workshop}:
  {Melody}}}.\hskip 1em plus 0.5em minus 0.4em\relax Hal Leonard Corporation,
  Jan. 2003.

\bibitem{kidde_learning_2020}
G.~Kidde, \emph{Learning {Music} {Theory} with {Logic}, {Max}, and
  {Finale}}.\hskip 1em plus 0.5em minus 0.4em\relax Routledge, Feb. 2020.

\bibitem{hernandez-olivan_music_2023}
C.~Hernandez-Olivan and J.~R. Beltrán, ``Music {Composition} with {Deep}
  {Learning}: {A} {Review},'' in \emph{Advances in {Speech} and {Music}
  {Technology}: {Computational} {Aspects} and {Applications}}.\hskip 1em plus
  0.5em minus 0.4em\relax Springer International Publishing, 2023, pp. 25--50.

\bibitem{wang_pop909_2020}
Z.~Wang, K.~Chen, J.~Jiang, Y.~Zhang, M.~Xu, S.~Dai, X.~Gu, and G.~Xia,
  ``{POP909}: {A} {Pop}-song {Dataset} for {Music} {Arrangement}
  {Generation},'' Aug. 2020, arXiv:2008.07142.

\bibitem{Inc.}
M.~Inc., ``mubert.com,'' https://mubert.com/.

\bibitem{Riffusion}
Riffusion, ``www.riffusion.com,'' https://www.riffusion.com/.

\bibitem{Schneider2023MosaiTG}
F.~Schneider, Z.~Jin, and B.~Sch{\"o}lkopf, ``Mo{\^u}sai: Text-to-music
  generation with long-context latent diffusion,'' \emph{ArXiv}, vol.
  abs/2301.11757, 2023.

\bibitem{Zhu2023ERNIEMusicTM}
P.~F. Zhu, C.~Pang, S.~Wang, Y.~Chai, Y.~Sun, H.~Tian, and H.~Wu,
  ``Ernie-music: Text-to-waveform music generation with diffusion models,''
  \emph{ArXiv}, vol. abs/2302.04456, 2023.

\bibitem{Huang2022MuLanAJ}
Q.~Huang, A.~Jansen, J.~Lee, R.~Ganti, J.~Y. Li, and D.~P.~W. Ellis, ``Mulan: A
  joint embedding of music audio and natural language,'' in \emph{International
  Society for Music Information Retrieval Conference}, 2022.

\bibitem{Agostinelli2023MusicLMGM}
A.~Agostinelli, T.~I. Denk, Z.~Borsos, J.~Engel, M.~Verzetti, A.~Caillon,
  Q.~Huang, A.~Jansen, A.~Roberts, M.~Tagliasacchi, M.~Sharifi, N.~Zeghidour,
  and C.~H. Frank, ``Musiclm: Generating music from text,'' \emph{ArXiv}, vol.
  abs/2301.11325, 2023.

\bibitem{NEURIPS2023_94b472a1}
J.~Copet, F.~Kreuk, I.~Gat, T.~Remez, D.~Kant, G.~Synnaeve, Y.~Adi, and
  A.~Defossez, ``Simple and controllable music generation,'' in \emph{Advances
  in Neural Information Processing Systems}, A.~Oh, T.~Naumann, A.~Globerson,
  K.~Saenko, M.~Hardt, and S.~Levine, Eds., vol.~36.\hskip 1em plus 0.5em minus
  0.4em\relax Curran Associates, Inc., 2023, pp. 47\,704--47\,720.

\bibitem{Eck2002}
D.~Eck and J.~Schmidhuber, ``Finding temporal structure in music: blues
  improvisation with lstm recurrent networks,'' in \emph{Proceedings of the
  12th IEEE Workshop on Neural Networks for Signal Processing}, 2002, pp.
  747--756.

\bibitem{Waite2016}
E.~Waite, ``Generating long-term structure in songs and stories,''
  magenta.tensorflow.org,
  https://magenta.tensorflow.org/2016/07/15/lookback-rnn-attention-rnn, Jul.
  2016.

\bibitem{Gillick2019LearningTG}
J.~Gillick, A.~Roberts, J.~Engel, D.~Eck, and D.~Bamman, ``Learning to groove
  with inverse sequence transformations,'' in \emph{International Conference on
  Machine Learning}, 2019.

\bibitem{Yang2017MidiNetAC}
L.-C. Yang, S.-Y. Chou, and Y.-H. Yang, ``Midinet: A convolutional generative
  adversarial network for symbolic-domain music generation,'' in
  \emph{International Societyfor Music Information Retrieval Conference}, 2017.

\bibitem{Trieu2018JazzGANI}
N.~Trieu and R.~M. Keller, ``Jazzgan : Improvising with generative adversarial
  networks,'' in \emph{International Workshop on Musical Metacreation}, 2018.

\bibitem{Dong2017MuseGANMS}
H.-W. Dong, W.-Y. Hsiao, L.-C. Yang, and Y.-H. Yang, ``Musegan: Multi-track
  sequential generative adversarial networks for symbolic music generation and
  accompaniment,'' in \emph{AAAI Conference on Artificial Intelligence}, 2017.

\bibitem{waite_generating_2016}
E.~Waite, ``Generating long-term structure in songs and stories,'' \emph{Web
  blog post. Magenta}, vol.~15, no.~4, 2016.

\bibitem{huang_pop_2020}
Y.~S. Huang and Y.~H. Yang, ``Pop {Music} {Transformer}: {Beat}-based
  {Modeling} and {Generation} of {Expressive} {Pop} {Piano} {Compositions},''
  in \emph{Proceedings of the 28th {ACM} {International} {Conference} on
  {Multimedia}}, ser. {MM} '20, Oct. 2020, pp. 1180--1188.

\bibitem{liu_symphony_2022}
J.~Liu, Y.~Dong, Z.~Cheng, X.~Zhang, X.~Li, F.~Yu, and M.~Sun, ``Symphony
  generation with permutation invariant language model,'' in
  \emph{International Society for Music Information Retrieval Conference},
  2022.

\bibitem{zhang_sdmuse_2023}
C.~Zhang, Y.~Ren, K.~Zhang, and S.~Yan, ``{SDMuse}: {Stochastic} {Differential}
  {Music} {Editing} and {Generation} via {Hybrid} {Representation},''
  \emph{IEEE Transactions on Multimedia}, pp. 1--9, 2023.

\bibitem{davis-mohammad-2014-generating}
H.~Davis and S.~Mohammad, ``Generating music from literature,'' in
  \emph{Proceedings of the 3rd Workshop on Computational Linguistics for
  Literature ({CLFL})}, A.~Feldman, A.~Kazantseva, and S.~Szpakowicz,
  Eds.\hskip 1em plus 0.5em minus 0.4em\relax Gothenburg, Sweden: Association
  for Computational Linguistics, Apr. 2014, pp. 1--10.

\bibitem{Rangarajan2015}
R.~Rangarajan, ``Generating music from natural language text,'' in \emph{2015
  Tenth International Conference on Digital Information Management (ICDIM)},
  2015, pp. 85--88.

\bibitem{zhang-etal-2020-butter}
Y.~Zhang, Z.~Wang, D.~Wang, and G.~Xia, ``{BUTTER}: A representation learning
  framework for bi-directional music-sentence retrieval and generation,'' in
  \emph{Proceedings of the 1st Workshop on NLP for Music and Audio (NLP4MusA)},
  S.~Oramas, L.~Espinosa-Anke, E.~Epure, R.~Jones, M.~Sordo, M.~Quadrana, and
  K.~Watanabe, Eds.\hskip 1em plus 0.5em minus 0.4em\relax Online: Association
  for Computational Linguistics, 16 Oct. 2020, pp. 54--58.

\bibitem{Wu2023}
S.~Wu and M.~Sun, ``Exploring the efficacy of pre-trained checkpoints in
  text-to-music generation task,'' in \emph{Creative AI Across Modalities
  workshop at AAAI 2023}, 11 2023.

\bibitem{Lu2023}
P.~Lu, X.~Xu, C.~W. Kang, B.~Yu, C.~Xing, X.~Tan, and J.~Bian, ``Musecoco:
  Generating symbolic music from text,'' \emph{ArXiv}, vol. abs/2306.00110,
  2023.

\bibitem{openai2024gpt4}
OpenAI, ``Gpt-4 technical report,'' OpenAI, Tech. Rep., 2024.

\bibitem{bubeck_sparks_2023}
S.~Bubeck, V.~Chandrasekaran, R.~Eldan, J.~A. Gehrke, E.~Horvitz, E.~Kamar,
  P.~Lee, Y.~T. Lee, Y.-F. Li, S.~M. Lundberg, H.~Nori, H.~Palangi, M.~T.
  Ribeiro, and Y.~Zhang, ``Sparks of artificial general intelligence: Early
  experiments with gpt-4,'' \emph{ArXiv}, vol. abs/2303.12712, 2023.

\bibitem{Hankun2006}
S.~Hankun, \emph{Melody Writing Tutorial (in Chinese)}.\hskip 1em plus 0.5em
  minus 0.4em\relax Xiamen University Press, 2006.

\bibitem{schoenberg_fundamentals_nodate}
A.~Schoenberg, \emph{\BIBforeignlanguage{en}{Fundamentals of musical
  composition}}, S.~Gerald and S.~Leonard, Eds.\hskip 1em plus 0.5em minus
  0.4em\relax Faber {$\&$} Faber, 1982.

\bibitem{mcnamee_review_1986}
A.~K. McNamee, ``Review of {Tonal} {Harmony}, with an {Introduction} to
  {Twentieth}-{Century} {Music},'' \emph{Journal of Music Theory}, vol.~30,
  no.~2, pp. 309--314, 1986.

\bibitem{Meng2022}
R.~Meng, C.~Zheng, X.~Li, J.~Sang, J.~Cai, J.~Wang, and X.~Wang, ``Emotionbox:
  A music-element-driven emotional music generation system based on music
  psychology,'' \emph{Frontiers in Psychology}, vol.~13, 08 2022.

\bibitem{Heinlein1928}
C.~P. Heinlein, ``The affective characters of the major and minor modes in
  music.'' \emph{Journal of Comparative Psychology}, vol.~8, pp. 101--142,
  1928.

\bibitem{Rigg1940}
M.~G. Rigg, ``Speed as a determiner of musical mood.'' \emph{Journal of
  Experimental Psychology}, vol.~27, pp. 566--571, 1940.

\bibitem{Thayer1990}
R.~E. Thayer, \emph{{The Biopsychology of Mood and Arousal}}.\hskip 1em plus
  0.5em minus 0.4em\relax Oxford University Press, 09 1990.

\bibitem{Forero2023Are}
J.~Forero, G.~Bernardes, and M.~Mendes, ``Are words enough? on the semantic
  conditioning of affective music generation.'' \emph{AIMC 2023}, aug 29 2023.

\bibitem{Livingstone2010}
S.~Livingstone, R.~Muhlberger, A.~Brown, and W.~Thompson, ``Changing musical
  emotion: A computational rule system for modifying score and performance,''
  \emph{Computer Music Journal}, vol.~34, pp. 41--64, 03 2010.

\bibitem{Alswaidan2020}
N.~Alswaidan and M.~Menai, ``A survey of state-of-the-art approaches for
  emotion recognition in text,'' \emph{Knowledge and Information Systems},
  vol.~62, 08 2020.

\bibitem{Deng2023}
J.~Deng and F.~Ren, ``A survey of textual emotion recognition and its
  challenges,'' \emph{IEEE Transactions on Affective Computing}, vol.~14,
  no.~1, pp. 49--67, 2023.

\bibitem{Mendes2023}
G.~A. Mendes and B.~Martins, ``Quantifying valence and arousal in text
  with multilingual pre-trained transformers,'' in \emph{Advances in
  Information Retrieval}, J.~Kamps, L.~Goeuriot, F.~Crestani, M.~Maistro,
  H.~Joho, B.~Davis, C.~Gurrin, U.~Kruschwitz, and A.~Caputo, Eds.\hskip 1em
  plus 0.5em minus 0.4em\relax Cham: Springer Nature Switzerland, 2023, pp.
  84--100.

\bibitem{walshaw_abcnotation_1995}
C.~Walshaw, ``abcnotation.com,'' abcnotation.com, 1995.

\bibitem{Devlin2019BERTPO}
J.~Devlin, M.-W. Chang, K.~Lee, and K.~Toutanova, ``Bert: Pre-training of deep
  bidirectional transformers for language understanding,'' in \emph{North
  American Chapter of the Association for Computational Linguistics}, 2019.

\bibitem{likert_technique_1932}
R.~Likert, ``A technique for the measurement of attitudes,'' \emph{Archives of
  Psychology}, vol. 22 140, pp. 55--55, 1932.

\end{thebibliography}

\end{document}